\documentclass[pdflatex,sn-mathphys-ay]{sn-jnl}

\usepackage{graphicx}%
\usepackage{multirow}%
\usepackage{amsmath,amssymb,amsfonts}%
\usepackage{amsthm}%
\usepackage{mathtools} 
\usepackage{mathrsfs}%
\usepackage[title]{appendix}%
\usepackage{xcolor}%
\usepackage{textcomp}%
\usepackage{manyfoot}%
\usepackage{booktabs}%
\usepackage{algorithm}%
\usepackage{algorithmicx}%
\usepackage{algpseudocode}%
\usepackage{listings}%
\hypersetup{
    colorlinks = true,
    urlcolor   = blue,
    citecolor  = black,
}
\usepackage{natbib}

\theoremstyle{thmstyleone}%
\theoremstyle{thmstyletwo}%

\theoremstyle{thmstylethree}%

\begin{document}

\title[Data-driven modeling of hypersonic flows in chemical non-equilibrium with catalytic surfaces]{Data-driven modeling of hypersonic flows in chemical non-equilibrium with catalytic surfaces}


\author*[1,2]{\fnm{Konstantinos} \sur{Sarras}}\email{konstantinos.sarras@lecnam.net}

\author[3]{\fnm{Louis} \sur{Walpot}}\email{louis.walpot@esa.int}

\author[2]{\fnm{Thierry} \sur{Magin}}\email{thierry.magin@vki.ac.be}

\author[4]{\fnm{Peter J.} \sur{Schmid}}\email{peter.schmid@kaust.edu.sa}

\author[1]{\fnm{Taraneh} \sur{Sayadi}}\email{taraneh.sayadi@cnam.fr}

\affil[1]{\orgdiv{MACS Laboratory}, \orgname{Conservatoire National des Arts et Métiers (CNAM)}, \orgaddress{ \city{Paris}, \postcode{75003}, 
\country{France}}}

\affil[2]{\orgname{von Karman Institute for Fluid Dynamics (VKI)}, \orgaddress{\street{Waterloosesteenweg 72}, \city{Sint-Genesius-Rode}, \postcode{1640}, \country{Belgium}}}

\affil[3]{\orgdiv{European Space Agency (ESA)}, \orgname{ESTEC}, \orgaddress{ \city{Noordwijk}, \postcode{2200 AG}, \country{The Netherlands}}}

\affil[4]{\orgdiv{Mechanical Engineering}, \orgname{King Abdullah University
of Science and Technology (KAUST)}, \orgaddress{ \city{Thuwal}, \postcode{23955}, \country{Saudi Arabia}}}


\abstract{ Hypersonic flows involve extreme thermochemical non-equilibrium, where strong energy dissipation leads to tightly coupled chemical reactions, radiation, and energy exchange. In this regime, surface chemistry, particularly catalytic wall reactions, can significantly affect boundary-layer composition and surface heat transfer. Accurate simulations of such flows may require repeated evaluations of detailed thermochemical libraries, which represent a major computational bottleneck in high-fidelity reactive-flow simulations. To mitigate this cost, we employ the data-driven reduced-order framework introduced by \cite{scherding2023data}, which combines nonlinear dimensionality reduction, community clustering, and local surrogate models to efficiently approximate high-dimensional thermochemical mappings. In this work, this framework is extended for the first time to hypersonic reactive flows with localized catalytic patches, whose rapid transitions in surface catalycity introduce sharp variations in wall chemistry and heat transfer. To address the increased complexity of the thermochemical state space, the dimensionality reduction method is enhanced with a Sammon-type stress penalty that mitigates topological folding of the latent manifold and improves the robustness of the clustering and surrogate stages. The resulting model accurately captures the effects of localized changes in surface catalycity, including sharp gradients in wall species mass fractions, diffusion fluxes, and surface heat transfer, while reducing the overall simulation cost by $50\%$ once the surrogate is constructed, without compromising accuracy.
}




\maketitle
\section{Introduction}\label{sec1}

The design of thermal protection systems (TPS) for atmospheric re-entry vehicles requires accurate prediction of the severe aerothermal environment generated during hypersonic flight \citep{andersonbook,gnoffo1999,bertin1994}. As a vehicle re-enters the atmosphere at speeds of several kilometers per second, a strong bow shock converts a large fraction of the freestream kinetic energy into thermal energy, raising post-shock temperatures to several thousand Kelvin. At these conditions, molecular nitrogen and oxygen undergo dissociation and exchange reactions produce nitric oxide, driving the gas far from thermal and chemical equilibrium \citep{candler2019,park1990,vincentiKruger1965}. When chemical reactions occur on time scales comparable to the characteristic flow time, chemical non-equilibrium persists throughout the shock layer and in the boundary layer developing along the vehicle surface. The associated aerodynamic heat transfer rates that reach tens of megawatts per square meter at the stagnation point \citep{fayRiddell1958,lees1956} impose extreme thermal loads on the TPS and drive the need for reliable, high-fidelity simulation tools capable of capturing the coupled aerothermochemical phenomena at play.

Apart from chemical reactions in the gas phase, the chemical reactivity of the vehicle's surface is an equally important mechanism for TPS design. Within the boundary layer, partially dissociated atomic species (predominantly atomic oxygen) may undergo heterogeneous catalytic recombination at the wall, releasing chemical energy that can substantially increase the wall heat flux relative to a fully non-catalytic surface \citep{goulard}. The magnitude of this catalytic contribution depends on the recombination probability of the TPS material and may vary by several orders of magnitude between candidate surface materials \citep{scott1980catalytic,nasuti1996material}. In practical re-entry configurations, neighboring tiles or shingles may exhibit significantly different catalytic properties, leading to abrupt spatial variations in surface species concentrations and wall heat flux across catalytic transitions \citep{baronets1991overequilibrium,aleksei}. To study the temperature variations induced by such catalytic discontinuities, ESA's Intermediate eXperimental Vehicle (IXV) \citep{ixv} flew a windward thermal protection system that combined ceramic-matrix-composite tiles of different catalytic behavior, including a dedicated highly-catalytic patch designed to produce a measurable temperature jump at the material interface \citep{ceglia2016experimental,viladegutCATE}. The present study is directly motivated by the need to model such catalytic transitions.

Numerical simulation has become an indispensable tool for the analysis and design of hypersonic flight systems. Early computational studies often relied on simplified gas models that neglected or strongly simplified thermochemical nonequilibrium effects \citep{Malik1989,Malik1990,Chang1997}. As understanding of hypersonic flows advanced, it became increasingly evident that calorically perfect or chemical equilibrium assumptions are inadequate for many practical applications \citep{candler2019,Josyula2015}. Consequently, numerical models have progressively evolved to incorporate increasingly realistic descriptions of high-temperature gas physics, including finite-rate chemical kinetics, thermal nonequilibrium, multicomponent transport and gas-surface interaction \citep{Miro2017,Wartemann2018,Hudson1997,Johnson2005, bellaspaper}. Today, state-of-the-art continuum simulations account for these effects by solving the compressible multi-species Navier-Stokes equations coupled with species conservation equations and finite-rate chemical source terms, while thermodynamic properties, transport coefficients, and chemical production rates are obtained from detailed thermochemical models \citep{Wright1998,Josyula2015,Candler2015US3D}. These evaluations are typically delegated to specialised thermochemical libraries, of which Mutation$^{++}$ \citep{mutation1}, EGLIB \citep{eglib2004}, PEGASE \citep{bottin1999}, and CHEMKIN \citep{kee2000} are representative examples. Such libraries provide accurate descriptions of high-temperature gas mixtures by accounting for complex chemical kinetics, transport phenomena, and thermodynamic properties over a broad range of thermochemical states. Their modular design also enables straightforward integration into a variety of computational fluid dynamics solvers, making them standard tools for high-fidelity hypersonic simulations.

The flexibility and physical fidelity of these libraries, however, come at a substantial computational cost. Since thermodynamic properties, transport coefficients, and reaction rates must be evaluated repeatedly at every grid point and every time step, thermochemical library calls often become the dominant computational bottleneck in large-scale reactive flow simulations. As demonstrated by \citet{scherding2023data} in the context of the Mutation$^{++}$ library, finite-rate reactive simulations can require more than an order of magnitude greater computational effort than their calorically perfect counterparts, with thermochemical evaluations accounting for the majority of the overall execution time. This rapidly increasing computational burden limits the applicability of high-fidelity simulations to large-scale, long-time, or parametric studies and motivates the development of efficient reduced-order strategies capable of accelerating thermochemical property evaluations without sacrificing predictive accuracy.

To circumvent this bottleneck, most high-performance solvers resort to hard-coded chemistry implementations \citep{direnzo2020}, which sacrifice generality and require significant re-engineering whenever the gas model changes. On the other hand, several strategies have been proposed to accelerate thermochemical closure while retaining the generality of a library-based approach. Structured look-up table (LuT) methods, in which library output is pre-tabulated and retrieved by interpolation at run time, have been applied to spray combustion \citep{franzelli2013}, real-gas turbomachinery flows \citep{pini2015} and hypersonic boundary layers under chemical equilibrium \citep{marxen2011}; their memory and construction costs grow exponentially with the input dimension, however, precluding direct use for the six-dimensional input spaces typical of five-species air mixtures. Pope's \textit{in situ} adaptive tabulation (ISAT) \citep{pope1997} addresses the storage problem by building the table adaptively during the simulation, but performs best when the accessed region of state space is compact and repeatedly revisited. In combustion, principal component analysis has also been used to reduce the number of transported equations, by solving for a few principal components instead of the species and parameterizing the thermochemistry in terms of these components \citep{sutherland2009,munoz2026}. More general surrogate modeling approaches including neural networks \citep{cybenko1989}, radial basis function (RBF) networks \citep{broomhead1988,buhmann2000} and kriging \citep{sacks1989, kleijnen2009} offer greater flexibility, but their direct application to high-dimensional thermochemical mappings is challenged by the curse of dimensionality and by the existence of sharp gradients and quasi-discontinuous transitions between thermochemical regimes, which degrade the accuracy of global surrogates \citep{bouhlel2016, bettebghor2011}. 

These limitations were addressed by \citet{scherding2023data,scherding2025} through the Reduced Order Nonlinear Approximation with Active Learning Procedure (RONAALP), a data-driven framework that compresses the high-dimensional thermochemical input space onto a low-dimensional manifold via nonlinear dimensionality reduction, partitions the manifold into physically consistent communities and fits local  surrogate models within each community. Unlike PCA-based transport models, the governing equations are left unchanged: the dimensionality reduction only organizes the input space of the surrogates, which replace the thermochemical closure. Applied to a Mach-10 adiabatic boundary layer and a shock-wave boundary-layer interaction, the framework achieved up to $70\%$ CPU time savings with no perceptible loss of accuracy in closed-loop \textit{a posteriori} assessments. More recently, RONAALP has also been employed to construct reduced thermochemical models for turbulent reactive jets through an \textit{a priori} flamelet-based formulation \citep{niemietz}, further demonstrating the flexibility of the approach across different reacting-flow regimes. Nevertheless, the application of RONAALP has so far been restricted to chemically inert walls, leaving the impact of surface chemistry unexplored. An extension of this framework to catalytic wall configurations is the focus of the present work.

In this paper, we specifically extend the RONAALP framework to hypersonic reactive flows over surfaces with heterogeneous catalytic effects. We consider a configuration featuring a localized catalytic patch, whose boundaries introduce rapid streamwise variations in surface properties. Including catalytic boundary conditions introduces additional thermochemical states associated with surface reaction balance, as well as implicitly coupling wall species concentrations to wall temperature. This significantly enlarges the thermochemical state space and increases the complexity of the surrogate modeling task. Following \cite{scherding2023data}, the present study focuses on the offline stage of RONAALP. The surrogate is trained on converged solutions and then used in closed loop on the same configuration. The objective is to establish whether this stage can represent the thermochemical state space enlarged by catalytic boundary conditions, and whether it remains stable when coupled to the catalytic surface balance across a rapid catalytic transition. Prediction of unseen configurations relies on the active-learning stage of the algorithm \citep{scherding2025}. It is not addressed here and is discussed in Sect.~\ref{sec:conclusions}.

The remainder of this paper is organized as follows. Section \ref{sec:configuration} describes the flow configuration under study. Section \ref{sec:methodology} presents the governing equations, together with the thermochemical models used both in the gas-phase and on the wall. Section \ref{sec:numerical} details the numerical framework used to solve the governing equations, including the flow solver and the full- and reduced-order thermochemical closures. Section \ref{sec:results} presents the results, including the full-order catalytic and non-catalytic boundary-layer solutions, the construction of the corresponding reduced-order models, and an assessment of their \textit{a priori} and \textit{a posteriori} performance. Finally, Section \ref{sec:conclusions} summarizes the main findings and outlines future perspectives.

\section{Flow configuration}\label{sec:configuration}

\begin{figure}[!t]
    \centering
    \includegraphics[scale=1,trim=0 0 0 0,clip]{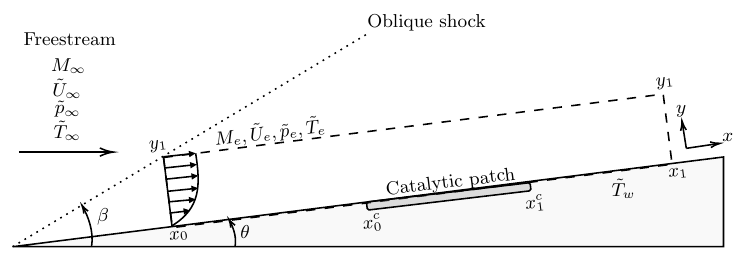} 
    \caption{Configuration of the hypersonic boundary-layer flow considered in the present study. An oblique shock wave of angle $\beta$, generated by a wedge of angle $\theta$, compresses a freestream flow characterized by $(M_\infty,\tilde{U}_\infty,\tilde{p}_\infty, \tilde{T}_\infty)$. The computational domain, indicated by dashed lines and defined in the coordinate system $(x,y)$ with $x\in[x_0,x_1]$ and $y\in[0,y_1]$, lies entirely downstream of the shock and therefore corresponds to a post-shock boundary layer with edge conditions $(M_e,\tilde{U}_e,\tilde{p}_e,\tilde{T}_e)$. The wall ($y=0$) has a temperature $\tilde{T}_w$, maintained under radiative-equilibrium conditions. A localized patch, fully catalytic with respect to oxygen recombination, is imposed over $x \in [x_0^c, x_1^c]$, with the remaining wall being non-catalytic.}
    \label{fig:configuration}
\end{figure}

\begin{table}[!h]
\centering
\small
\setlength{\tabcolsep}{3pt}
\resizebox{\textwidth}{!}{%
\begin{minipage}{\textwidth}
\centering
\begin{tabular}{c c c c c | c c c c c c c  }

${M}_\infty$ & $\tilde{U}_\infty (\mathrm{m/s})$ & $\tilde{p}_\infty (\mathrm{Pa})$ &
$\tilde{T}_\infty (\mathrm{K})$ & $\tilde{\rho}_\infty (\mathrm{kg/m^3})$ &
$M_e$ & $\tilde{U}_e (\mathrm{m/s})$ & $\tilde{p}_e (\mathrm{Pa})$ &
$\tilde{T}_e (\mathrm{K})$ & $\tilde{\rho}_e (\mathrm{kg/m^3})$ &
$\theta (^\circ)$ &
$\beta (^\circ)$ \\
\hline
23.4 & 6863.1 & 21.96 & 214.39 & 3.57$\times10^{-4}$ &
5.75 & 6791 & 1782 & 3106 & 1.72$\times10^{-3}$ & 17 & 20.9 \\
\end{tabular}
\vspace{2mm}
\makebox[\linewidth][c]{%
\begin{tabular}{ c c c c c c c c c c c}

 $x_0$ & $x_1$ & $y_0$ & $y_1$ &
$x_0^c$ & $x_1^c$ & $N_x$ & $N_y$ & $\Delta x$ & $\Delta y_1$ & $\eta$ \\
\hline
 14 & 100 & 0 & 3.5 &
40 & 60 & 860 & 220 & 0.1 & 0.0012 & 0.15 \\
\end{tabular}
}
\end{minipage}
}
\caption{Freestream and post-shock boundary-layer edge conditions (top table), together with the geometric parameters and mesh resolution (bottom table) used for the hypersonic boundary-layer simulations. }
\label{tab:freestream_caseA_essential}
\end{table}

In this paper, we consider a hypersonic boundary-layer configuration representative of the flow over the windward surface of a re-entry vehicle downstream of a shock, with the gas in a dissociated state. Under the local approximation, the post-shock flow is idealized as a canonical two-dimensional boundary layer developing behind a planar oblique shock induced by a wedge geometry, as shown in Fig.~\ref{fig:configuration}. The problem is defined by the freestream conditions $({M}_\infty,\tilde{p}_\infty,\tilde{\rho}_\infty,\tilde{T}_\infty)$. 
Here, freestream conditions at $M_\infty = 23.4$ are considered, representative of the flight conditions of the ESA IXV mission \citep{viladegutCATE}.

The wedge, with deflection angle $\theta$, generates an oblique shock with shock angle $\beta$ measured relative to the freestream direction. The computational domain lies entirely downstream of the shock and therefore
consists of a post-shock boundary layer. Its edge Mach number $M_e$, pressure $\tilde p_e$ and temperature $\tilde T_e$ are obtained from the oblique-shock relations of a calorically perfect gas ($\gamma = 1.4$), assuming the weak-shock solution branch, while the edge
composition is set to chemical equilibrium at $(\tilde T_e, \tilde p_e)$ (Appendix~\ref{secA:edge}). The wedge angle is treated as a design parameter and is set to $\theta=17^\circ$, yielding both hypersonic post-shock conditions ($M_e=5.75$) and sufficiently high edge temperatures ($\tilde{T}_e \gtrsim 3000\,\mathrm{K}$) to promote significant dissociation of molecular oxygen \citep{andersonbook}. The resulting freestream and post-shock states are summarized in Table \ref{tab:freestream_caseA_essential}.

The computational domain is defined in the wall-bounded coordinate system $(x,y)$ and extends, in non-dimensional coordinates, over ${x} \in [x_0, x_1]$ and ${y} \in [0,y_1]$. The wall ($y=0$) is maintained at a temperature $\tilde{T}_w$ determined from radiative equilibrium. A highly emissive material with emissivity coefficient $\epsilon=0.9$ is considered, resulting in wall temperatures between $1200$ and $1500\, \mathrm{K}$, consistent with those encountered in thermal protection systems under hypersonic aerothermal loading \citep{Panerai2012}.

The baseline wall is assumed to be non-catalytic, while a patch that is fully catalytic with respect to oxygen recombination  is prescribed over the streamwise interval $x \in [x_0^c,x_1^c]$. The domain size is defined by the nondimensional parameters $x_0 = 14$, $x_1 = 100$, and $y_1 = 3.5$, while the catalytic patch is imposed between $x_0^{c} = 40$ and $x_1^{c} = 60$, with the non-dimensional coordinate $ x =\tilde{x}/ \tilde{L}_{\mathrm{ref}}$ being related to a reference length of $\tilde{L}_{\mathrm{ref}} = 0.0873 \, \mathrm{m}$. In terms of streamwise Reynolds number $ Re_x = {\tilde{\rho}_e \tilde{U}_e \tilde{x}}/{\tilde{\mu}_e}$, the computational domain corresponds to the interval $Re_x \in [1.4 \times 10^5, 1\times 10^6]$, while the catalytic patch occupies the subinterval $Re_x \in [4 \times 10^5, 6 \times 10^5]$.  Such $Re_x$ conditions fall within the range typically encountered in hypersonic boundary-layer studies of re-entry configurations \citep{fedorov2011transition}.

\section{Governing equations}\label{sec:methodology}


In this study, hypersonic reactive flows are modeled using the compressible multi-species Navier-Stokes equations. Considering a mixture composed of $\mathcal{S}$ species, these equations, in non-dimensional form, read
\begin{align}
\frac{\partial \rho}{\partial t}
+ \nabla \cdot (\rho \mathbf{u})
&= 0, \label{eqn:continuity} \\
\frac{\partial (\rho \mathbf{u})}{\partial t}
+
\nabla \cdot (\rho \mathbf{u} \otimes \mathbf{u})
&= - \nabla p + \nabla \cdot \boldsymbol{\tau}, \label{eqn:momentum} \\
\frac{\partial (\rho E)}{\partial t}
+
\nabla \cdot \left((\rho E + p)\mathbf{u}\right)
&=
\nabla \cdot (\boldsymbol{\tau}\cdot\mathbf{u})
-
\nabla \cdot \mathbf{q}, \label{eqn:energy}\\
\frac{\partial \rho Y_s  }{\partial t}
+ \nabla \cdot (\rho Y_s \mathbf{u} + \mathbf{J}_s )
&= \dot{\omega}_s,
\qquad \forall s \in \mathcal{S} \label{eqn:species}, 
\end{align}
comprising the classical compressible Navier-Stokes equations \eqref{eqn:continuity}-\eqref{eqn:energy}, augmented by a system of $\mathcal{S}$ mass conservation equations \eqref{eqn:species}, one for each species $s \in \mathcal{S}$. In the latter, $t$ denotes time, $\mathbf{u} = (u,v)^\top$ is the velocity vector, $p$ the thermodynamic pressure, $\rho$ the fluid's density and $E$ is the fluid's total specific energy, defined as $E = e + \frac{1}{2}\|\mathbf{u}\|^2$, with $e$ being the thermodynamic internal energy of the mixture. In the species equation,  $Y_s=\rho_s/\rho$ denotes the species mass fractions satisfying $\sum_{s \in \mathcal{S}} Y_s = 1$, with $\rho_s$ corresponding to the partial density of species $s$, $\mathbf{J}_s$ is the diffusion flux, accounting for the diffusion of species due to concentration gradients and $\dot{\omega}_s$ corresponds to the species net chemical mass production rates due to reactions. Details on the definition of $\mathbf{J}_s$ and $\dot{\omega}_s$ are given in Appendix~\ref{appendix:thermochemical-kinetics}.

In this context, the mixture's density $\rho$, internal energy $e$, and enthalpy $h$ follow from the species properties as
\begin{equation}
\rho = \sum_{s \in \mathcal{S}} \rho_s, \quad e = \sum_{s \in \mathcal{S}} Y_s e_s, \quad h = \sum_{s \in \mathcal{S}} Y_s h_s
\end{equation}
while the viscous stress tensor $\boldsymbol{\tau}$ and the heat flux vector $\mathbf{q}$ are given by
\begin{align}
\boldsymbol{\tau}
=
\mu \left(
\nabla \mathbf{u}
+
\nabla \mathbf{u}^\top
-
\frac{2}{3}
(\nabla \cdot \mathbf{u}) \mathbf{I}
\right), \quad
\mathbf{q}
=
-\frac{\kappa}{\mathrm{Re}_e \mathrm{Pr}_e}
\nabla T + \sum_{s \in \mathcal{S}} \mathbf{J}_s h_s,
\end{align}
where $\mu$ is the dynamic viscosity, $\mathbf{I}$ is the identity tensor and $\kappa$ the thermal conductivity, with the  heat flux accounting for both Fourier heat conduction and energy transport due to species diffusion.  

All aerodynamic and thermochemical quantities are non-dimensionalized using boundary-layer-edge reference quantities, denoted by the subscript $(\cdot)_e$, according to
\begin{equation}
\begin{gathered}
\boldsymbol{x} = \frac{\tilde{\boldsymbol{x}}}{\tilde{L}_{\mathrm{ref}}}, \quad
t = \frac{\tilde{t}\,\tilde{a}_{e}}{\tilde{L}_{\mathrm{ref}}}, \quad
\rho_s = \frac{\tilde{\rho}_s}{\tilde{\rho}_e}, \quad
p = \frac{\tilde{p}}{\tilde{\rho}_e \tilde{a}_e^2}, \quad
\mathbf{u} = \frac{\tilde{\mathbf{u}}}{\tilde{a}_e}, \quad
\kappa = \frac{\tilde{\kappa}}{\tilde{\kappa}_e}, \quad
\mu = \frac{\tilde{\mu}}{\tilde{\mu}_e}, \\
T = \frac{\tilde{T}}{(\gamma_e - 1)\tilde{T}_e}, \quad
e = \frac{\tilde{e}}{\tilde{a}_e^2}, \quad
h = \frac{\tilde{h}}{\tilde{a}_e^2}, \quad
\dot{\omega}_s =
\frac{\tilde{\dot{\omega}}_s}
{\tilde{\rho}_e \tilde{a}_e / \tilde{L}_{\mathrm{ref}}}, \quad
\mathbf{J}_s =
\frac{\tilde{\mathbf{J}}_s}
{\tilde{\rho}_e \tilde{a}_e},
\end{gathered}
\label{eq:nondimensional}
\end{equation}
where $\tilde{(\cdot)}$ denotes dimensional quantities,
$\tilde{L}_{\mathrm{ref}}$ is the dimensional reference length,
$\gamma_e $ is the isentropic exponent, $\tilde{a}_e = \sqrt{\gamma_e R_e \tilde{T}_e} $ is the speed of sound at the boundary-layer edge and $R_e$ denotes the specific gas constant evaluated at the edge. The dimensionless Mach, Reynolds and Prandtl numbers that globally define the problem are defined as 
\begin{equation}\label{eqn:RePrEc}
{M}_e = \frac{\tilde{U}_e}{\tilde{a}_e}, \quad
{Re}_e = \frac{\tilde{\rho}_e \tilde{a}_e \tilde{L}_{\mathrm{ref}}}{\tilde{\mu}_e}, \quad
{Pr}_e = \frac{\tilde{\mu}_e \tilde{c}_{p,e}}{\tilde{\kappa}_e}. 
\end{equation}

\subsection{Thermochemical model in the gas-phase and on the wall}\label{sec:finiterate}

Throughout this study, the flow is assumed to be chemically reactive, with reaction time scales comparable to those of the flow, i.e., in chemical nonequilibrium. The gas is further assumed to be in thermal equilibrium, so that the translational, rotational, vibrational and electronic energy modes of all species are described by a single temperature.

A five-species atmospheric mixture, $\mathcal{S} = \{\text{N}_2, \text{O}_2, \text{N}, \text{O}, \text{NO}\}$, is considered. In the gas phase, species undergo five reversible reactions following Park’s five-reaction mechanism for dissociated air \citep{parkmodel}:
\begin{equation}
\begin{rcases}
\mathrm{R}_1 &: \text{N}_2 + \text{M} \;\rightleftharpoons\; 2\text{N} + \text{M}, \\
\mathrm{R}_2 &: \text{O}_2 + \text{M} \;\rightleftharpoons\; 2\text{O} + \text{M}, \\
\mathrm{R}_3 &: \text{NO} + \text{M} \;\rightleftharpoons\; \text{N} + \text{O} + \text{M}, \\
\mathrm{R}_4 &: \text{N}_2 + \text{O} \;\rightleftharpoons\; \text{NO} + \text{N}, \\
\mathrm{R}_5 &: \text{NO} + \text{O} \;\rightleftharpoons\; \text{N} + \text{O}_2
\end{rcases}
\end{equation}
where $\text{M}$ denotes a third body, representing any of the five species. Gas-phase chemistry is modeled using finite-rate kinetics based on the law of mass action. Species production and consumption rates are obtained from reaction stoichiometry together with the corresponding forward and backward reaction rates, thereby accounting for both dissociation--recombination and exchange processes under nonequilibrium conditions. Details on the chemical source terms, including the Arrhenius rate expressions, are provided in Appendix~\ref{appendix:thermochemical-kinetics}.

Here, solid walls are also assumed to be chemically reactive through catalytic surface reactions. One heterogeneous reaction is considered, corresponding to recombination of atomic oxygen into its diatomic form, or
\begin{equation}\label{eqn:recombination}
\mathrm{R}_{w} : \text{O} + \text{O} + w \;\rightarrow\; \text{O}_2 + w,
\end{equation}
where $w$ denotes the wall acting as a catalyst. Other heterogeneous reactions such as nitrogen recombination ($\mathrm{N}+\mathrm{N}+w\rightarrow\mathrm{N_2}+w$) or the formation of nitric oxide ($\mathrm{N}+\mathrm{O}+w\rightarrow
\mathrm{NO}+w$), are not considered. Under the conditions of this study, nitrogen is essentially molecular at the boundary-layer edge ($Y_{\mathrm{N},e} \approx 10^{-4}$, against $Y_{\mathrm{O},e} \approx 0.18$, Appendix~\ref{secA:edge}), so that these reactions are limited by a flux of nitrogen atoms about three orders of magnitude smaller than that of oxygen atoms, and their contribution to the wall composition and heat flux is negligible.

Catalytic wall boundary conditions are formulated by requiring conservation of species mass and total energy at the gas--surface interface \citep{bellaspaper,baskaya}. Under the assumption that surface chemistry reaches equilibrium on a time scale much shorter than that of the flow, the wall fluxes satisfy the following quasi-steady relations,
\begin{align} \mathbf{J}_s \cdot \boldsymbol{\hat{n}}_w &= \dot{\omega}_{w,s}, \quad \forall s \in \mathcal{S}, \label{eqn:smb} \\ 
\bigg(\frac{\kappa_w}{{Re}_e {Pr}_e }\nabla T - \sum_{s \in \mathcal{S}} \mathbf{J}_s h_s \bigg)\cdot \boldsymbol{\hat{n}}_w &= \sigma \epsilon T_w^4. \label{eqn:seb}\end{align}
The first expression, Eq.~\eqref{eqn:smb}, states that the diffusive flux of each species at the surface is balanced by its net rate of production (or consumption) due to catalytic reactions. Eq.~\eqref{eqn:seb} enforces the thermal balance at the wall by accounting for the combined contributions of conductive heat transfer and enthalpy transport by species diffusion, balanced by radiative heat emission. In these equations, $\boldsymbol{\hat{n}}_w$ denotes the outward unit normal vector, $T_w$ the wall temperature, $\kappa_w$ is the thermal conductivity at the wall and $\dot{\omega}_{w,s}$ the net surface reaction rate of species $s$. The parameter $0\leq \epsilon \leq 1$ denotes the wall emissivity, while $\sigma = {\tilde{\sigma}\,\big[(\gamma_e -1)\tilde{T}_e\big]^{4}}/{\tilde{\rho}_e\,\tilde{a}_e^{3}}$ is the nondimensional Stefan-Boltzmann constant, with $\tilde{\sigma}$ being the dimensional Stefan--Boltzmann constant.

The catalytic behavior of the wall is represented at the macroscopic scale as a stochastic surface process \citep{goulard}, characterized by a recombination probability (or catalytic efficiency) $\gamma_\mathrm{O}$ for the atomic oxygen species, defined as
\begin{equation}
\gamma_\mathrm{O} = \frac{\mathcal{F}_{\mathrm{O},\mathrm{reac}}}{\mathcal{F}_{\mathrm{O},\mathrm{imp}}},
\end{equation}
where $\mathcal{F}_{\mathrm{O},\mathrm{reac}}$ denotes the flux of oxygen atoms that undergo recombination at the surface, while $\mathcal{F}_{\mathrm{O},\mathrm{imp}}$ is the incoming flux of particles striking the wall. The parameter satisfies $0 \leq \gamma_\mathrm{O} \leq 1$ and is prescribed in the model, with the limiting cases $\gamma_\mathrm{O} = 1$ and $\gamma_\mathrm{O} = 0$
corresponding to a wall that is fully catalytic with respect to oxygen recombination and to a completely inert wall, respectively. In the following,
``fully catalytic'' refers to $\gamma_\mathrm{O} = 1$ in this restricted sense. Assuming that the gas adjacent to the wall follows a Maxwellian velocity distribution at the wall temperature, the impinging flux can be expressed as
\begin{equation}
\mathcal{F}_{\mathrm{O},\mathrm{imp}} = n_\mathrm{O} \sqrt{\frac{k_B T_w}{2 \pi m_\mathrm{O}}},
\end{equation}
where $k_B$ is the Boltzmann constant, $m_\mathrm{O}$ is the mass and $n_\mathrm{O}=\rho_\mathrm{O}/m_\mathrm{O}$ the number density of the oxygen species. Given a prescribed value of $\gamma_\mathrm{O}$, the corresponding wall source terms for atomic and molecular oxygen are given by
\begin{equation}
\dot{\omega}_{w,\mathrm{O}} = -\gamma_\mathrm{O}  m_\mathrm{O}  \mathcal{F}_{\mathrm{O},\mathrm{imp}}, \qquad
\dot{\omega}_{w,\mathrm{O}_2} = \gamma_\mathrm{O}  m_\mathrm{O}  \mathcal{F}_{\mathrm{O},\mathrm{imp}}, 
\label{eqn:wallsource} 
\end{equation}
whereas wall source terms of all non-reacting species vanish,
$
\dot{\omega}_{w,s}=0, \,  s \neq \mathrm{O},\mathrm{O}_2
$.

\section{Numerical framework}\label{sec:numerical}
In this work, simulations are performed using the Multi-Species hyperSonic solver (MS$^2$) \citep{margaritisetal}, built upon a high-fidelity compressible Navier–Stokes solver for non-reacting flows \citep{nagarajan1,sayadijfm}. 
The governing equations are spatially discretized using finite-difference schemes on a structured, staggered grid. A uniform mesh with $N_x=860$ points is employed in the streamwise direction, yielding a streamwise grid spacing $\Delta x=(x_1 - x_0 )/(N_x - 1)=0.1$, while $N_y=220$ points are used in the wall-normal direction with grid clustering near the wall according to
$ y(m)=y_0+(y_1-y_0)\left[(1-\eta)\zeta^3+\eta\zeta\right], $
where $m \in [1,N_y]$, $\zeta=(m-1)/(N_y-1)$, and $\eta$ is the stretching parameter. A value of $\eta=0.15$ yields first cell-centre-to-wall spacing of $\Delta y_1=y(2)/2=0.0012$. With this grid, the boundary layer is resolved by 78 to 105 points across its thickness $\delta_{99}$ over the computational domain, with $\Delta y_1/\delta_{99} < 0.004$, and the near-wall layer in which the composition is modified by the catalytic patch is resolved by at least 58 points in the wall-normal direction. The same spatial discretization and wall-normal grid stretching were assessed in the grid-convergence study of \citet{margaritisetal} for a Mach-10 reactive hypersonic boundary layer.

The implementation of the catalytic boundary conditions (Eqs.~\eqref{eqn:smb}–\eqref{eqn:seb}) in MS$^2$ is described in Appendix~\ref{app:catalyticBC}, along with its verification. To model the catalytic patch, the recombination coefficient $\gamma_\mathrm{O}$ is varied abruptly at its boundaries. A step change in $\gamma_\mathrm{O}$ cannot be handled by the high-order, non-dissipative discretization of the solver: the resulting discontinuity in the wall boundary condition produces spurious oscillations near the patch edges and prevents the convergence of the coupled surface balance equations. For this reason, $\gamma_\mathrm{O}$ is defined as a continuous function of $x$, which varies rapidly but smoothly across the patch boundaries, based on Gaussian-like exponential terms, or 
\begin{equation} \gamma_\mathrm{O}(x) = \begin{cases} 0, & x < x^c_0 \text{ or } x > x^c_1,\\[2mm] \Gamma_{\mathrm{O}} \cdot \Bigl( 1 - e^{- \frac{(x - x^c_0)^2}{4 \delta^2}} \Bigr) \Bigl( 1 - e^{- \frac{(x - x^c_1)^2}{4 \delta^2}} \Bigr), & x^c_0 \le x \le x^c_1, \end{cases}  \label{eqn:gammafunction} \end{equation}
where $0\leq \Gamma_\mathrm{O}\leq 1$ is the maximum value attained by $\gamma_\mathrm{O}$ and $\delta$ controls the smoothness of the transition at the patch boundaries $x_0^c$ and $x_1^c$. The recombination coefficient therefore satisfies $\gamma_\mathrm{O}=0$ outside the catalytic patch and reaches $\gamma_\mathrm{O}=\Gamma_\mathrm{O}$ within the patch. The value $\delta = 6\Delta x$, which ensures numerical stability, is adopted throughout the present study.

The complete set of configuration parameters and mesh resolutions is summarized in the bottom row of Table \ref{tab:freestream_caseA_essential}. Further details regarding the numerical flow solver, including the spatial discretization, time integration schemes and boundary condition implementation can be found in \cite{margaritisetal}. To enforce global mass conservation, the continuity equation \eqref{eqn:continuity} is solved together with the species conservation equations \eqref{eqn:species} for all species but one, forming the set of transported species $\mathcal{S}_\mathrm{t}\subset\mathcal{S}$, with $|\mathcal{S}_\mathrm{t}|=|\mathcal{S}|-1$. The density of the remaining species is recovered from the closure relation $\rho=\sum_{s\in\mathcal{S}}\rho_s$. The omitted species is typically chosen as the most chemically inert constituent of the mixture.

\subsection{Thermochemical closure -- Mutation$^{++}$}\label{sec:mutationclosure}

The governing equations \eqref{eqn:continuity}-\eqref{eqn:species} require closure relations to determine the thermochemical properties of the gas mixture, including species thermodynamic quantities, transport coefficients, and chemical source terms. In hypersonic flows, the high temperatures encountered across shock waves and within boundary layers promote dissociation, recombination, and other nonequilibrium phenomena, making these properties strongly dependent on the local thermochemical state. Accurate evaluation of such quantities is therefore essential for reliable flow predictions. In this work, thermochemical effects are incorporated through coupling with Mutation$^{++}$ \citep{mutation1}, an open-source library developed at the von Karman Institute for Fluid Dynamics (VKI) which supplies consistent models for transport, thermodynamic, and chemical-kinetic properties of multicomponent reacting gas mixtures.

The coupling is formulated as a pointwise input–output relation $\mathbf{z} = f(\mathbf{x})$, where $f : \mathbb{R}^D \to \mathbb{R}^{D_z}$ denotes the nonlinear mapping that collects all thermochemical evaluations provided by Mutation$^{++}$, from the input space $\mathbb{R}^D$ onto the output space $\mathbb{R}^{D_z}$. The input vector consists of the multi-component flow state vector
$\mathbf{x} = (\rho, \rho_s, \rho e)^\top, \, s \in\mathcal{S}_\mathrm{t}$, while the output vector $\mathbf{z} = (\mu, \kappa,p, T, D_s,  h_s, \dot{\omega}_s)^\top, \, s \in\mathcal{S}$ gathers all thermochemical quantities required for closure. These include transport properties ($\mu, \kappa, D_s$), thermodynamic variables ($p, T, h_s$) and chemical source terms ($\dot{\omega}_s$) in the gas phase.

As demonstrated by \cite{scherding2023data}, Mutation$^{++}$ introduces a substantial computational overhead in hypersonic reactive flow simulations, with this increased cost arising from the repeated evaluation of complex nonlinear thermochemical relations at every computational point and iteration. Consequently, the cost of real-gas simulations is significantly larger than that of equivalent chemically inert computations. To mitigate this, \cite{scherding2023data} proposed a reduced-order modeling framework. The main objective of the algorithm is to replace the expensive thermochemical evaluations performed by Mutation$^{++}$ with an accurate surrogate model (referred to here as Mutation$^{++}$ Light) constructed from a reduced set of representative thermochemical states. In the present work, an updated and modified version of this reduced-order framework is used to accelerate the evaluation of the thermochemical closure, while maintaining the fidelity required for high-speed reacting flow simulations. The main components of the present algorithm are briefly summarized in the following subsections.

\begin{figure}[!t]
    \centering
    \includegraphics[scale=1,trim=0 0 0 0,clip]{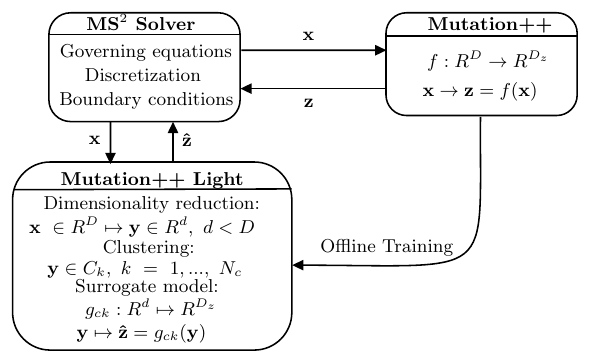} 
    \caption{ General schematic representation of the RONAALP algorithm. The original nonlinear thermochemical closure $\mathbf{z}=f(\mathbf{x})$ is replaced by the reduced-order surrogate model Mutation$^{++}$ Light, which combines dimensionality reduction, clustering, and local surrogate approximations to infer thermochemical quantities $\hat{\mathbf{z}}$ from the flow state vector $\mathbf{x}$. The surrogate model is constructed during an offline training phase using data generated by the full Mutation$^{++}$ library. Figure is adapted from \cite{scherding2023data}.}
    \label{fig:ronaalp}
\end{figure}

\subsection{Reduced thermochemical closure -- Mutation$^{++}$ Light}

The Reduced Order Nonlinear Approximation with Active Learning Procedure (RONAALP) algorithm of \cite{scherding2023data,scherding2025} was specifically designed for accelerating simulations of hypersonic non-equilibrium flows.

The original RONAALP algorithm consists of three consecutive stages, illustrated schematically in Fig.~\ref{fig:ronaalp}. First, the high-dimensional thermochemical input space is projected onto a reduced latent manifold using a nonlinear dimensionality reduction technique. The resulting latent representation is then partitioned into several physically meaningful regions using a community detection algorithm. Finally, independent surrogate models are constructed within each region to locally approximate the nonlinear thermochemical mapping. In the present work, the nonlinear dimensionality reduction method is extended to explicitly penalize topological folding of the latent manifold. These three steps form the offline stage of RONAALP. In its complete form \citep{scherding2025}, the algorithm also includes an online active-learning stage. During the simulation, states outside the training
range are detected with an extrapolation metric in the latent space, evaluated with the full library, and incorporated in the RBF networks through online $k$-means and a Schur-complement update. In this stage the encoder and the community partition are kept fixed. As shown by \cite{scherding2025}, this stage is required whenever the simulation visits states outside the training data, for instance in unsteady flows or at different flow conditions. In the present work only the offline stage is used.

We stress that the surface mass and energy balances are governing equations of the problem, in the same way as Eqs.~\eqref{eqn:continuity}-\eqref{eqn:species}. They are therefore solved in every simulation, with or without the surrogate. The reduced-order model replaces neither them nor the Navier-Stokes equations. It replaces only the pointwise closure $\mathbf{z}={f}(\mathbf{x})$.

\subsubsection{Topology-preserving dimensionality reduction}\label{sec:IO-E}

Considering $N$ realizations of the thermochemical closure function $f$, let the input matrix be $ \mathbf{X} = (\mathbf{x}_1,\ldots,\mathbf{x}_N)^\top,$ and the corresponding output matrix be
$ \mathbf{Z} = (\mathbf{z}_1,\ldots,\mathbf{z}_N)^\top$. The first step of the algorithm aims to compress the high-dimensional input
state $\mathbf{X}\in \mathbb{R}^{N \times D}$ into a lower-dimensional latent
representation $\mathbf{Y}\in \mathbb{R}^{N \times d}$ with $d<D$ while retaining the
information relevant to the prediction of the thermochemical quantities.
Dimensionality reduction here is based on an input/output encoder (IO-E). As opposed to a classical auto-encoder which learns a latent representation $\mathbf{Y}$ through an encoder--decoder architecture trained to minimize the input reconstruction error, the IO-E is trained directly on the thermochemical output $\mathbf{Z} \in \mathbb{R}^{N \times D_z}$, with the objective of approximating the nonlinear mapping $\mathbf{Z}=f(\mathbf{X})$. The main objective of the IO-E is therefore to minimize the following functional
\begin{equation}
    \mathcal{L}_{\mathrm{recon}} = \frac{1}{N_b D_z}\sum_{i=1}^{N_b}
    \left\lVert f(\mathbf{X}_i)-\hat{\mathbf{Z}}_i \right\rVert^2 ,
\end{equation}
where during training, the loss is evaluated over randomly sampled mini-batches of size $N_b$, yielding a stochastic estimate of the full-dataset objective. Here $\hat{\mathbf{Z}}_i$ denotes the network prediction of the target thermochemical quantities for the $i$-th sample in the current mini-batch. Similar to a conventional auto-encoder, the IO-E architecture contains an encoder ($\mathbb{R}^D \to \mathbb{R}^d$) and a decoder ($\mathbb{R}^d \to \mathbb{R}^{D_z}$). Nevertheless, only the encoder is retained in the final model to provide the reduced latent coordinates $\mathbf{Y}$.

Owing to the increased complexity of the present problem, which comprises multiple physically distinct regions associated with different thermochemical states, the latent manifold is more susceptible to topological distortions during nonlinear dimensionality reduction. To prevent the nonlinear encoder from excessively distorting the
neighborhood structure of the input space, in this paper a stress penalty inspired by the classical Sammon mapping \citep{sammon1969nonlinear} is added to the standard reconstruction loss. As a result,
the final objective functional $\mathcal{L}$ to be minimized combines the
output reconstruction error with a relative-distance-preservation term,
\begin{equation}
    \mathcal{L} = \mathcal{L}_{\mathrm{recon}}
    \;+\; \lambda_{\mathrm{stress}}
    \underbrace{\frac{1}{N_b^2}\sum_{i=1}^{N_b}\sum_{j=1}^{N_b}
    \left( \frac{d^{Y}_{ij}}{\bar{d}^{Y}}
    - \frac{d^{X}_{ij}}{\bar{d}^{X}} \right)^2}_{\mathcal{L}_{\mathrm{stress}}},
    \label{eq:ioe_loss}
\end{equation}
where $d^{X}_{ij} = \lVert \mathbf{X}_i - \mathbf{X}_j \rVert$ and
$d^{Y}_{ij} = \lVert \mathbf{Y}_i - \mathbf{Y}_j \rVert$ define the
pairwise Euclidean distances in the input space and latent space
respectively, both restricted to the current mini-batch, while $\bar{d}^{X}$ and $\bar{d}^{Y}$ denote their respective means averaged over all ordered pairs within the mini-batch. Normalizing each distance by its batch-wise mean makes $\mathcal{L}_{\mathrm{stress}}$ scale-invariant, so it penalizes relative rather than absolute distortions of the pairwise-distance structure between the input and latent representations. 

The weight $\lambda_{\mathrm{stress}}\geq 0$ controls the trade-off between reconstruction accuracy and preservation of the input space's structure in the latent embedding ($\lambda_{\mathrm{stress}}=0$ recovers the unconstrained IO-E while for $\lambda_{\mathrm{stress}}\gg1$ the mapping approaches a near-isometric embedding of the input space, analogous to a classical Sammon mapping). In practice, $\lambda_{\mathrm{stress}}$ is selected so as to remove the
topological folding of the latent manifold, which makes the community partition reproducible (Appendix~\ref{app:topology}).

\subsubsection{Community clustering}\label{sec:newmann}

Once the reduced latent manifold $\mathbf{Y}$ has been obtained, the data is partitioned into $N_c$ communities, $\mathbf{Y} \in C_k, \, k=1,\ldots,N_c$,
corresponding to regions that exhibit similar thermochemical behavior. Since the relationship between the flow variables and the thermochemical quantities is highly nonlinear and may vary significantly across the state space, constructing a single global surrogate can lead to reduced accuracy. The identification of distinct communities allows the subsequent surrogate models to focus on a narrower range of physical conditions and therefore achieve improved predictive performance.

The partitioning of the latent space is performed using Newman's spectral community detection algorithm \citep{newman-reference}. A graph representation of the latent manifold is first constructed by connecting neighboring samples based on their Euclidean distance. The corresponding distance matrix is binarized using a connectivity threshold $\ell$, defined as a fraction of the mean Euclidean distance in the latent space, $\ell = c_\ell\, \bar{d}^Y$, where $c_{\ell}>0$ is a scaling parameter and $\bar{d}^{Y}$ denotes the average pairwise distance between latent samples. Newman's algorithm is then applied to identify densely connected regions of the graph, yielding $N_c$ communities without requiring the number of clusters to be specified \textit{a priori}. This feature is particularly advantageous for complex thermochemical datasets, where the number of distinct regimes is generally unknown.

The community detection is performed on a representative subsample, $N_s$, of the training data to reduce the computational cost associated with graph construction. Once the communities have been identified, a random forest classifier \citep{breiman2001random} is trained using the latent coordinates as input and the assigned community labels as outputs, using the entropy criterion for node splitting. It is subsequently employed to propagate the community labels to the full training dataset and to classify new samples during the inference stage.

\subsubsection{Local surrogate modeling}\label{sec:RBF}
The final component of the RONAALP algorithm consists of constructing local surrogate models for each identified community $C_k$. As in \cite{scherding2023data}, radial basis function (RBF) networks are employed here due to their favorable compromise between accuracy, computational efficiency, and straightforward training procedure. In addition, RBF networks support adaptive strategies such as sequential learning as well as growth of the network structure, making them suitable for reduced-order modeling applications.

For each community $C_k$, an independent RBF network is trained to approximate the local nonlinear mapping between the latent coordinates $\mathbf{Y}$ and the thermochemical outputs $\mathbf{Z}$, containing a hidden layer composed of $N_R$ radial basis functions characterized by centers $\mathbf{y}_i^c \in C_k$ and corresponding weights $a_i$. To guarantee a well-posed interpolation problem, the RBF sum is augmented with a low-order polynomial term, so that the resulting approximation reads
\begin{equation}
g_{ck}(\mathbf{y}) = \sum_{i=1}^{N_R} a_i \, \phi\left(\lVert \mathbf{y}-\mathbf{y}_i^c \rVert\right)
\;+\; \sum_{j=1}^{N_p} c_j \, p_j(\mathbf{y}),
\label{eq:rbf_interpolant}
\end{equation}
where $\phi$ denotes the selected radial basis kernel ($\phi(r) = r^2 \log r$ in this work) and $\{p_j\}_{j=1}^{N_p}$ is a basis of polynomials up to a prescribed degree (here third degree), added to ensure conditional positive-definiteness of the interpolation system, with $N_p = 20$ in this study. The locations of the RBF centers $\mathbf{y}_i^c$  are determined through k-means clustering of the latent coordinates in each community $C_k$ whereas the optimal coefficients $a_i$ and $c_j$ are obtained by solving the corresponding linear system directly rather than relying on iterative optimization algorithms. To ensure continuity of the surrogate across community boundaries, the nearest centroids shared between adjacent communities are included in the training set. 

\section{Results}\label{sec:results}

\subsection{Catalytic and non-catalytic boundary layer solutions}

All simulations are run until a time-independent, laminar state is reached. The effect of surface catalycity is investigated by comparing the two limiting cases of the catalytic model: a fully non-catalytic wall ($\Gamma_{\mathrm{O}}=0$) and a wall with a fully catalytic patch ($\Gamma_{\mathrm{O}}=1$) with respect to oxygen recombination.

\begin{figure}[!t]
\centering
\begin{tabular}{ll}
 (a) & (b) \\
\includegraphics[scale=0.43,trim=0 5 0 0,clip]{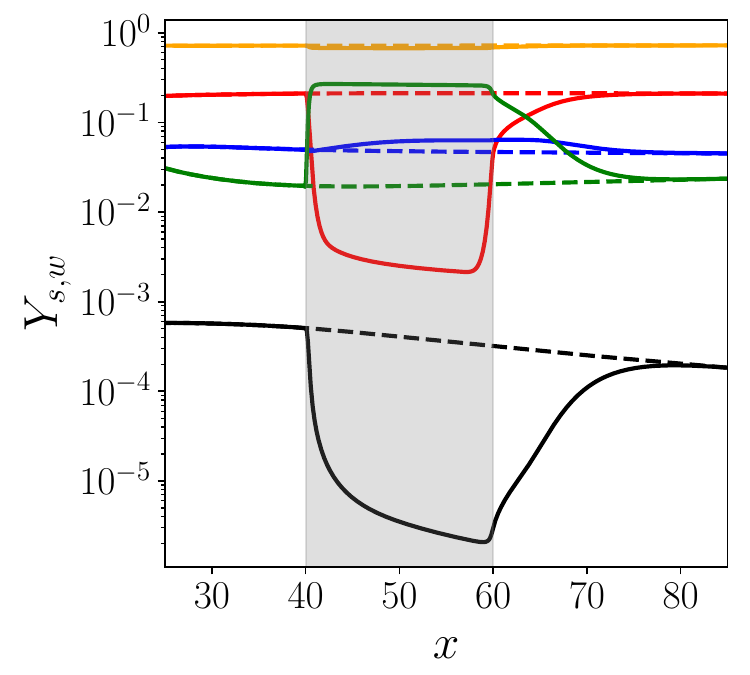} & 
\includegraphics[scale=0.43,trim=0 5 0 0,clip]{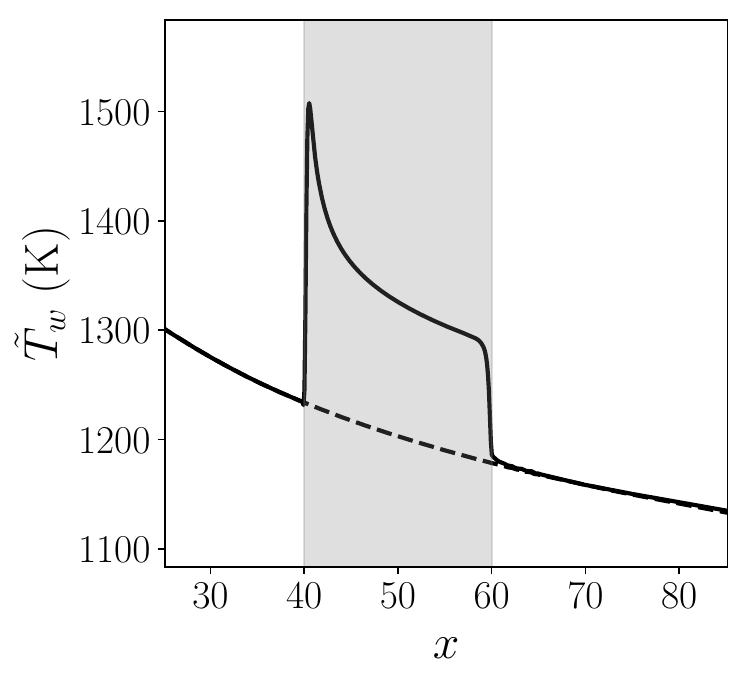} \\
 (c) & (d) \\
\includegraphics[scale=0.43,trim=0 0 0 8,clip]{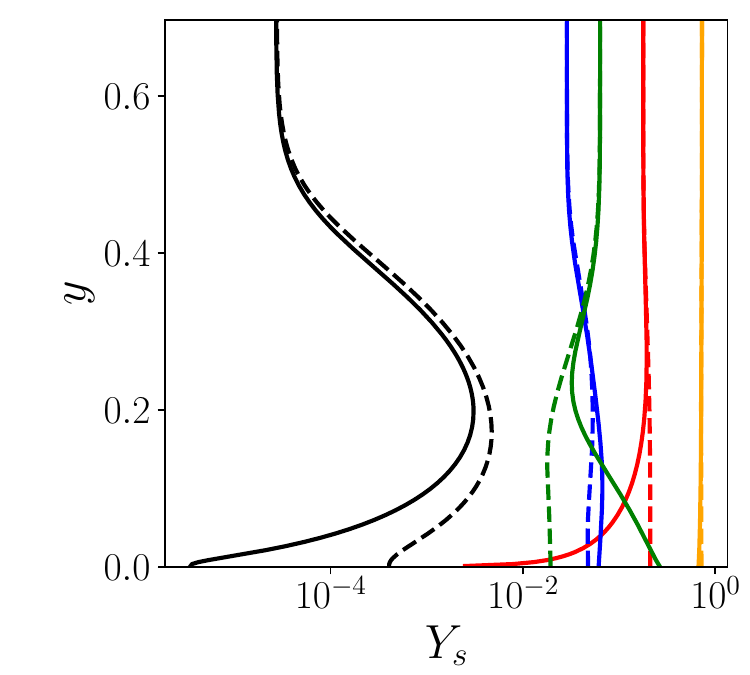} & 
\includegraphics[scale=0.43,trim=0 0 0 8,clip]{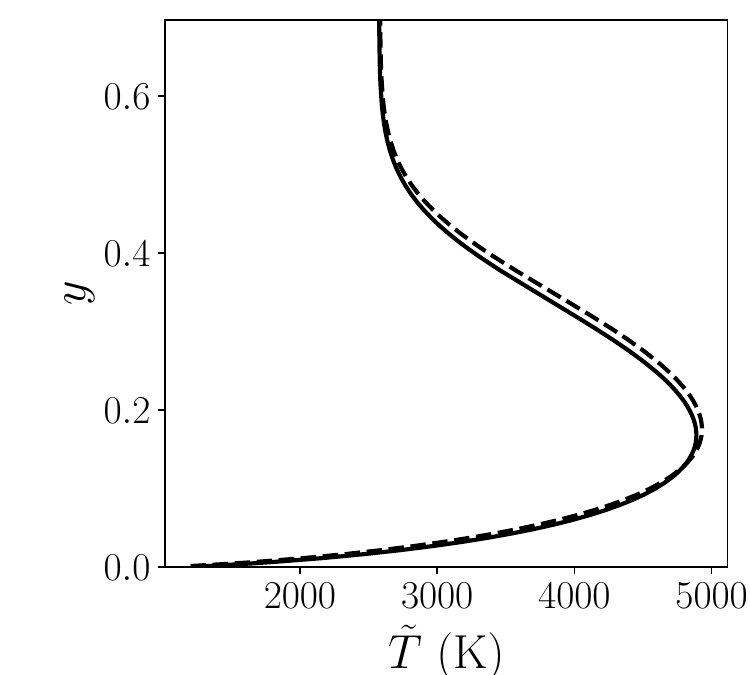} 
\end{tabular}
\caption{Steady-state solution of the $M_e=5.75$ boundary-layer flow. (a) Species mass fractions at the wall $Y_{s,w}=\rho_{s,w}/\rho_w$ (in logarithmic scale) and (b) dimensional wall temperature $\tilde{T}_w$ as a function of the nondimensional streamwise coordinate $x$. Profiles of (c) species mass fractions $Y_s$ (in logarithmic scale) and (d) dimensional temperature $\tilde{T}$ as function of the nondimensional distance to the wall $y$, extracted at $x=50$. Colors in (a,c) correspond to $\mathrm{N}$ (black), $\mathrm{O}$ (red), $\mathrm{NO}$ (blue), $\mathrm{N_2}$ (yellow) and $\mathrm{O_2}$ (green). The grey shaded region on (a) and (b) indicates the extent of the catalytic patch. Solid lines denote results obtained with the fully-catalytic patch ($\Gamma_\mathrm{O}=1$), whereas dashed lines correspond to the non-catalytic-wall case ($\Gamma_\mathrm{O}=0$).}
\label{fig:steadystate}
\end{figure}

Figure \ref{fig:steadystate} shows key characteristics of the steady-state boundary-layer flow, with and without catalytic wall effects. We first examine the species concentrations. Figures \ref{fig:steadystate}(a) and (c) show, respectively, the species mass fractions at the wall, $Y_{s,w}=\rho_{s,w}/\rho_w$, and the wall-normal profiles of species mass fractions $Y_s$ as a function of the nondimensional wall distance $y$, with the latter being extracted at the streamwise location corresponding to the midpoint of the catalytic patch ($x=50$). The most pronounced effect of the catalytic patch on species concentrations is the strong depletion of atomic oxygen $\mathrm{O}$, accompanied by a substantial enrichment of molecular oxygen $\mathrm{O_2}$, which follows directly from the imposed surface recombination reaction (Eq.~\eqref{eqn:recombination}). A noticeable reduction in both atomic and molecular nitrogen, $\mathrm{N}$ and $\mathrm{N_2}$, is also observed, followed by an increase of $\mathrm{NO}$, despite the absence of direct nitrogen-wall reactions. This behavior arises indirectly through multicomponent diffusion and species coupling: the strong redistribution of oxygen species modifies the local mixture composition and, through the mass-fraction constraint $\sum_s Y_s = 1$, induces compensating variations in the remaining species. Note that although catalytic reactions are restricted to $x\in[40,60]$, their influence extends downstream of the patch, up to approximately $x\approx80$; species concentrations recover only gradually toward the non-catalytic profiles, due to the convective-diffusive relaxation timescale of the boundary layer.

The wall-normal profiles of $Y_s$ further reveal an important signature of catalytic activity; the formation of steep concentration gradients in the immediate vicinity of the wall. In particular, the reacting species exhibit non-zero wall-normal gradients, with $\partial Y_{\mathrm{O}}/\partial y|_{y=0}>0$ and $\partial Y_{\mathrm{O_2}}/\partial y|_{y=0}<0$, where the sign of each gradient follows directly from the corresponding value of the catalytic production term $\dot{\omega}_{w,s}$ through the surface mass balance (Eq.~\eqref{eqn:smb}), as the diffusive flux $\mathbf{J}_s$ scales with species mass-fraction gradients (Eq.~\eqref{eqn:ficklaw}). The effects of catalysis on species are confined to the boundary layer ($y\lesssim 0.5$), while the free-stream composition remains unchanged.

Species concentrations directly influence the thermodynamic state of the boundary layer. This effect is clearly illustrated in Fig.~\ref{fig:steadystate}(b), reporting the dimensional wall temperature $\tilde{T}_w$ as a function of $x$. Catalysis leads to a significant increase in wall temperature, driven by the exothermic nature of the surface recombination reactions, as accounted for in the surface energy balance (Eq.~\eqref{eqn:seb}). Within the catalytic patch, the wall temperature increases from approximately $1250$ to $1350 \, \mathrm{K}$, corresponding to a rise of about $100\, \mathrm{K}$. A prominent feature is the local temperature overshoot observed at the upstream transition point ($x=40$). This peak arises from the sudden exposure of excess atomic oxygen, convected from upstream non-catalytic conditions, to an active catalytic surface, which causes a narrow heat-release overshoot before the species adjust to the catalytic equilibrium. No analogous peak occurs at $x=60$, where the incoming flow is already oxygen-poor. Such overshoots are a known feature of catalytic transitions \citep{chung,ottens,aleksei,baronets1991overequilibrium}, manifesting as locally enhanced heating above equilibrium levels at catalytic junctions. Its amplitude and streamwise extent in the present simulations, however, depend on the width of the imposed transition in $\gamma_\mathrm{O}$, and are therefore not quantitatively predicted here. The effect of catalysis on temperature is predominantly localized near the wall. As shown in Fig.~\ref{fig:steadystate}(d), the wall-normal temperature profiles exhibit only minor differences between catalytic and non-catalytic cases away from the wall, indicating that, for the present configuration, catalysis does not significantly alter the overall thermal structure of the boundary layer. 

The discussion above confirms that catalysis alters the thermochemical state through changes in species composition, surface heat release, and coupled transport processes, and that this occurs in a spatially localized manner, particularly in the near-wall regions and around the catalytic patch. This behavior motivates an examination of whether the data-driven reduction framework can adapt to these spatial variations and correctly identify the physically relevant regions of the flow.

\subsubsection{Training data acquisition} 

The input $\mathbf{X}$ and output matrices $\mathbf{Z}$ used to train the reduced-order model are constructed by pointwise sampling of the converged time-independent flow fields for both the catalytic and non-catalytic cases. Rather than using all grid points, a subset of locations is selected through a non-uniform random sampling procedure with an increased density of samples in the vicinity of the wall. This strategy ensures that the reduced-order model accurately captures the sharp thermal and chemical gradients developing in the near-wall region, where the effects of catalytic surface reactions are most pronounced. This sampling procedure yields a total of $N=43\,900$ training points per configuration, which are split randomly
into a $N_{\mathrm{train}}$ training set ($80\%$) and a $N_{\mathrm{test}}$ test set ($20\%$). All stages of the model are built from the training set only, and the test set is used only to evaluate the prediction error.

Prior to training, input and output matrices are mapped to $[0,1]^{N \times D}$ and $[0,1]^{N \times D_z}$ using statistics
fitted on the training set only, or
\begin{equation}
{\mathbf{X}} =\frac{{\mathbf{X}}^{\textrm{r}}-\min({\mathbf{X}}^{\textrm{r}})}
{\max({\mathbf{X}}^{\textrm{r}})-\min({\mathbf{X}}^{\textrm{r}})},
\qquad
{\mathbf{Z}} = \frac{{\mathbf{Z}}^{\textrm{r}}-\min({\mathbf{Z}}^{\textrm{r}})}
{\max({\mathbf{Z}}^{\textrm{r}})-\min({\mathbf{Z}}^{\textrm{r}})},
\end{equation}
where ${\mathbf{X}}^{\textrm{r}}$ and ${\mathbf{Z}}^{\textrm{r}}$ denote the raw, unscaled input and output data, respectively, and the $\min(\cdot)$ and
$\max(\cdot)$ operators are applied component-wise over the training samples. The normalization is applied so that all features contribute on a comparable numerical scale during training, preventing variables with intrinsically larger magnitudes from dominating the optimization process and ensuring that the loss function assigns similar importance to all states.

\subsection{Catalytic and non-catalytic reduced order models}
\subsubsection{Latent space manifolds}

\begin{figure}[!t]
\centering
\begin{tabular}{ll}
(a) & 
(b) \\
\includegraphics[scale=0.6,trim=60 30 70 60,clip]{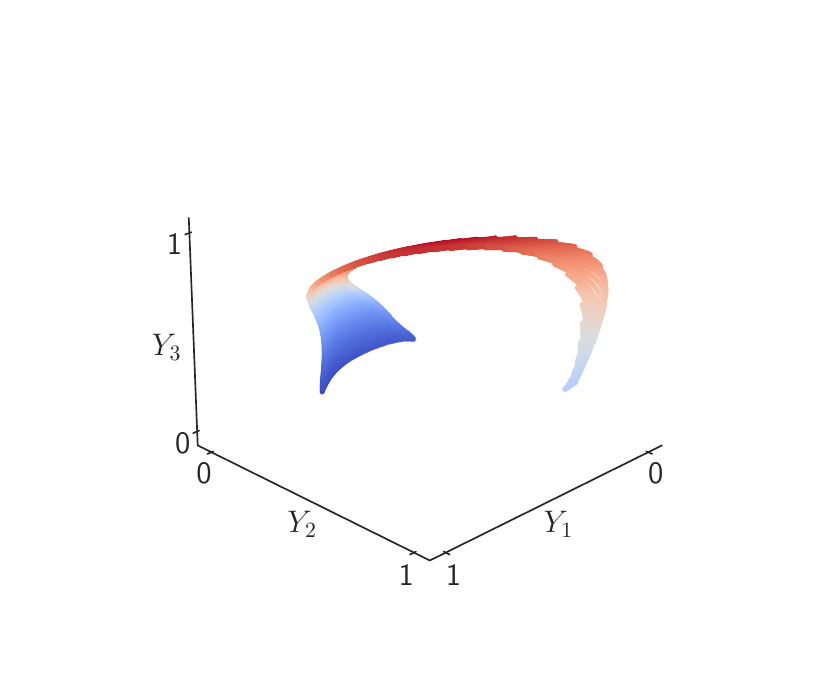} &
\hspace{-2em}
\includegraphics[scale=0.6,trim=40 30 0 60,clip]{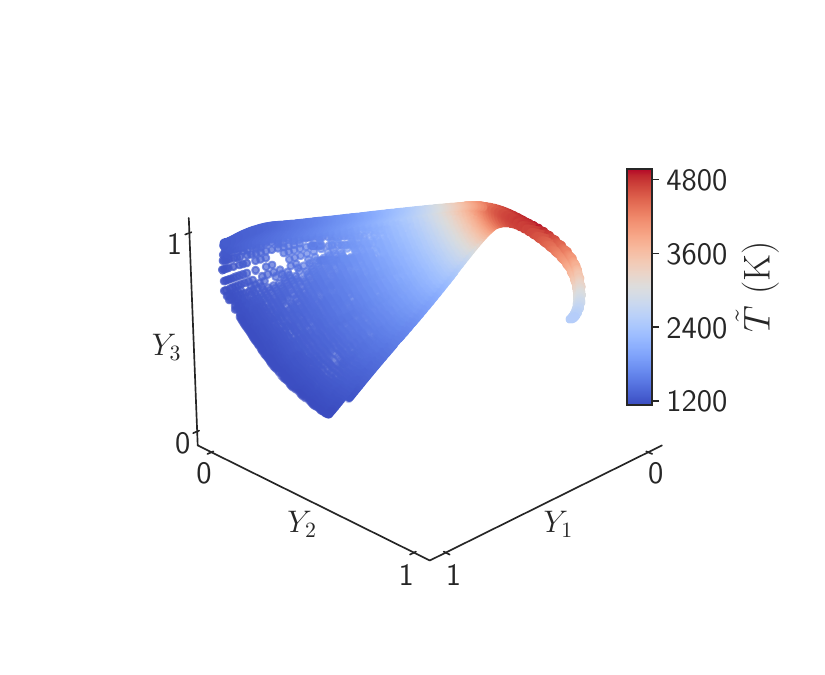}
\end{tabular}
\caption{Three-dimensional manifolds in latent space ($Y_1,Y_2,Y_3$), obtained by performing dimensionality reduction using the topology-preserving IO-E on the (a) non-catalytic  and (b) fully catalytic  solution datasets. The manifolds are coloured according to contours of the dimensional temperature field $\tilde{T}$.}
\label{fig:manifolds}
\end{figure}

As a first step, a low-dimensional latent-space representation of the problem is constructed using the nonlinear IO-E architecture, with the embedding constrained by the stress regularization term, as described in Sect.~\ref{sec:IO-E}. To inform the choice of latent-space dimensionality $d$, the intrinsic dimensionality of the thermochemical data is first examined by performing principal component analysis (PCA) on the input space ${\mathbf{X}}$. The optimal latent space dimension is determined using a criterion based on the cumulative variance $V(n)$. Specifically, we select the smallest $n$ such that $V(n)$ exceeds a $95\%$ threshold, $n^{\ast}  = \min\{n : V(n) \ge 0.95\}$, which by the Eckart–Young theorem is equivalent to bounding the relative linear reconstruction error of the input data by $1-V(n) \leq 5\%$. As detailed in Appendix~\ref{app:pca}, the six-dimensional input state of the non-catalytic data is well approximated by a two-dimensional linear subspace, whereas the catalytic data require three dimensions, indicating that catalysis increases the intrinsic dimensionality of the thermochemical state space. The latent dimension $d=3$ is adopted for both. Since this criterion concerns only inputs, the choice was confirmed by training the complete model for $d = 2$ to $5$ and comparing the output prediction errors (Appendix~\ref{app:pca}): $d=3$ is the smallest dimension that reaches the target accuracy for the catalytic data and the most accurate one for the non-catalytic data. The implementation details of the input-output encoder, including the network architecture and training hyper-parameters, are summarized in Appendix~\ref{app:ae_parameters}.

Figure \ref{fig:manifolds} shows the three-dimensional latent-space manifolds obtained for the (a) noncatalytic and (b) catalytic datasets, obtained for $\lambda_{\mathrm{stress}}=1$, which assigns equal importance to the output reconstruction error and the preservation of the latent-space topology (Appendix~\ref{app:topology}). To provide a physical interpretation of the latent representation, the manifolds are colored according to contours of the dimensional temperature field $\tilde{T}$, corresponding to one of the output variables of the closure function. In both cases, the resulting manifolds exhibit a smooth and continuous structure, indicating that the nonlinear dimensionality reduction successfully captures the underlying organization of the thermochemical state space. Regions characterized by different temperature levels, and therefore by distinct thermochemical conditions, are mapped onto well-defined regions of the latent space, while preserving the main physical gradients of the original high-dimensional space. However, the present manifolds display a distinct feature; specifically, two regions characterized by low-to-moderate temperatures ($\tilde{T} < 3600 \, \mathrm{K}$) are mapped onto separate areas of the latent space, with these regions being disconnected by an intermediate zone of high temperature ($\tilde{T}\gtrsim 4800 \, \mathrm{K}$). This behavior differs from that observed by \cite{scherding2023data} for boundary layers over adiabatic (hot) walls, where states with similar temperature levels were consistently mapped to the same regions of the latent space. As will be shown in the following subsection, these two regions, despite exhibiting similar temperature levels, correspond to distinct thermochemical states and are associated with physically different locations within the boundary layer.

\subsubsection{Community clusters}\label{sec:results-clusters}

\begin{figure}[t]
\begin{tabular}{ll}
(a) & (b) \\
\includegraphics[scale=0.6,trim=70 15 70 60,clip]{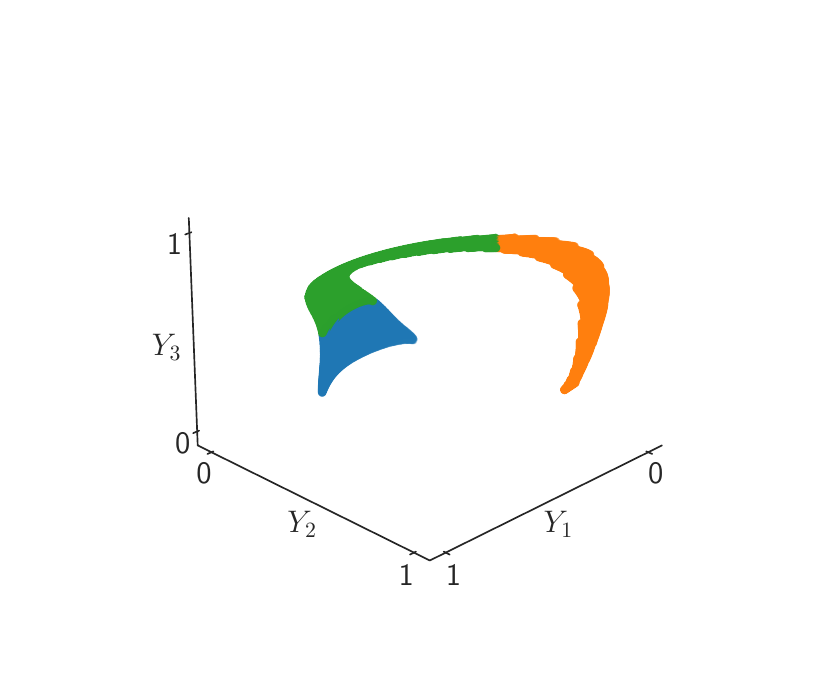} &
\raisebox{0.7em}{\includegraphics[scale=0.56,trim=20 10 0 10,clip]{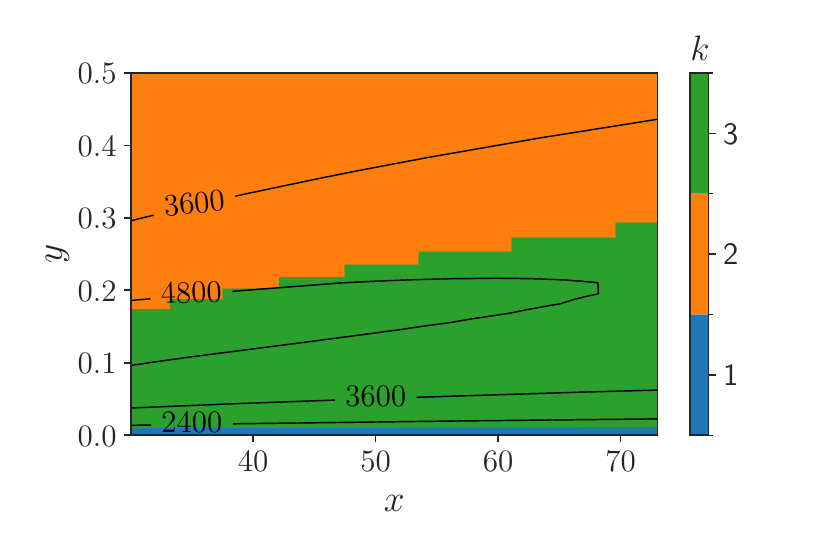}} \\[-1em]
(c) & (d) \\
\includegraphics[scale=0.6,trim=70 15 70 60,clip]{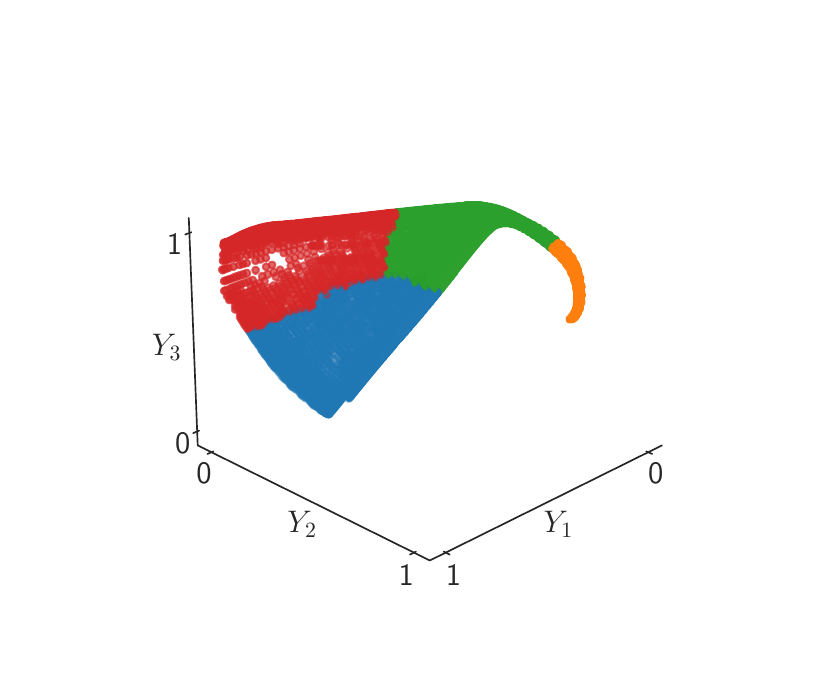} &
\raisebox{0.7em}{\includegraphics[scale=0.56,trim=20 10 0 10,clip]{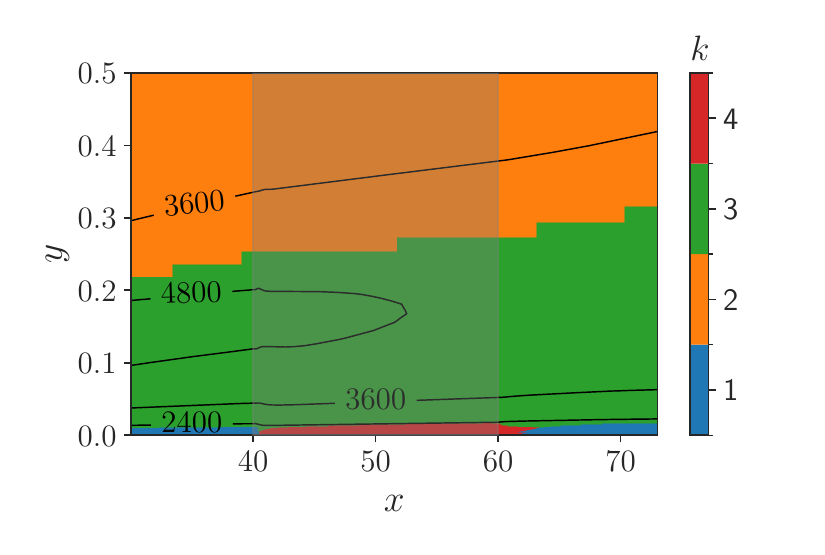}}
\end{tabular}
\caption{Community clusters labeled by their index $k$, identified using the Newman algorithm for the non-catalytic (a,b) and fully catalytic (c,d) datasets, shown in the (a,c) latent space $(Y_1,Y_2,Y_3)$ and the (b,d) physical space $(x,y)$. The isolines in (b,d) represent contours of the dimensional temperature field $\tilde{T}$. The grey shaded region in (d) indicates the extent of the catalytic patch. }
\label{fig:clusters}
\end{figure}

Following the construction of the low-dimensional manifold, the latent space is partitioned into clusters using the Newman community detection algorithm (Sect.~\ref{sec:newmann}). Figure \ref{fig:clusters} presents the communities identified by Newman's algorithm for both the non-catalytic and catalytic datasets, obtained from a random subsample of $N_s=10000$ points extracted from the respective training sets using a connectivity threshold factor of $c_{\ell}=0.8$. The resulting partition is shown in the low-dimensional latent space together with its projection onto the physical domain, allowing each community to be associated with a distinct thermochemical region of the boundary layer. For the non-catalytic configuration, the algorithm separates the manifold into three well-defined clusters. Cluster $k=1$ encompasses the cold fluid ($\tilde{T}\lesssim 2400 \, \mathrm{K}$) adjacent to the wall ($y \lesssim 0.01$), cluster $k=3$ corresponds to the high-temperature portion of the boundary layer, with temperatures reaching $\tilde{T}\gtrsim 4800 \, \mathrm{K}$, while cluster $k=2$ represents the outer freestream region ($y\gtrsim 0.25$), characterized by temperatures close to the prescribed edge value, $\tilde{T}\approx \tilde{T}_e=3106 \,\mathrm{K}$. This result shows that the latent coordinates encode sufficient thermochemical information to identify physically distinct regimes, despite the absence of spatial information during clustering, with each cluster mapping onto a spatially coherent region of the boundary layer.

In contrast, the catalytic configuration is partitioned into four communities. While clusters $k=3$ and $k=2$ remain associated with the high-temperature region and outer freestream, respectively, the near-wall region is divided into two distinct communities. Cluster $k=1$ corresponds to the non-catalytic wall, whereas cluster $k=4$ is confined to the catalytic patch ($40 < x < 60$) and persists slightly downstream of its trailing edge due to convective transport. The appearance of a fourth community demonstrates that the thermochemical modifications induced by surface recombination are sufficiently pronounced to form a distinct region of the latent manifold, which the community detection algorithm identifies as an independent cluster.

The robustness of these partitions was assessed by training the encoder three times with different random initializations and repeating the community detection, each time with a new random subsample and with $c_{\ell}$ ranging between $0.7$ and $1.3$. The number of communities $N_c$ is controlled by $c_{\ell}$ and decreases as the threshold increases. A community confined to the catalytic patch is found in all trainings for $0.7 \le c_{\ell} \le 1.0$ and in none for $c_{\ell} > 1.0$, where it
merges with the non-catalytic near-wall community. No such community is found for the non-catalytic dataset, whose partition contains three communities for all values of $c_{\ell}$. Concurrently, the subsample size has no significant effect on $N_c$; with $c_{\ell} = 0.8$, the same communities are obtained for $N_s = 2\,500$, $5\,000$ and $10\,000$, where increasing $N_s$ only increases the detection cost. The catalytic regime is therefore a robust feature of the latent manifold, but its identification as a separate community requires a sufficiently fine tuned threshold, $ 0.7 \le c_{\ell} \le 1.0$ for the present study.

\subsubsection{Surrogate model accuracy}

\begin{figure}
\begin{tabular}{ll}
(a) & (b) \\
{\includegraphics[scale=0.56,trim=20 10 10 10,clip]{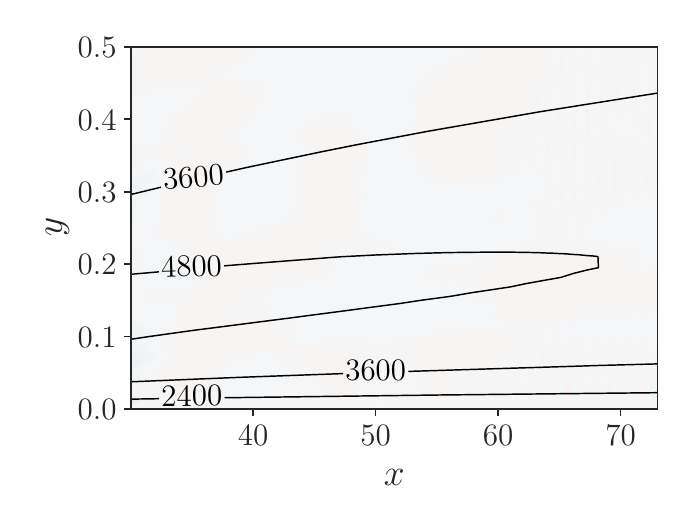}} &{\includegraphics[scale=0.56,trim=60 10 0 10,clip]{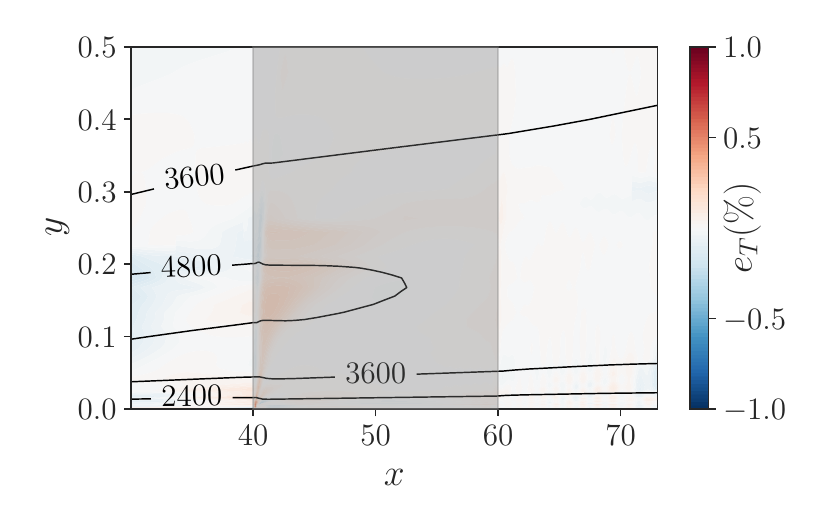} }
\end{tabular}
\caption{Relative temperature error field $e_T = (\hat{T}-T)/T \, (\%)$ for the (a) non catalytic and (b) catalytic patch cases. Isolines represent contours of the dimensional temperature field $\tilde{T}$ while the gray shaded region in (b) indicates the extent of the catalytic patch.}
\label{fig:modelerror}
\end{figure}

\begin{table}[t]
  \centering
  \setlength{\tabcolsep}{3pt}
  \begin{tabular}{lcccccccccccc}
     & $T$ & $\mu$ & $\kappa$ & $h_{\mathrm{O}}$ & $h_{\mathrm{O_2}}$ & $D_{\mathrm{N}}$ & $D_{\mathrm{O}}$ & $D_{\mathrm{O_2}}$
     & $\dot\omega_{\mathrm{N}}$ & $\dot\omega_{\mathrm{O}}$ & $\dot\omega_{\mathrm{NO}}$ & $\dot\omega_{\mathrm{O_2}}$ \\
    \hline
    $\overline{e}$ & 0.109 & 0.079 & 0.105 & 0.025 & 0.179 & 0.232 & 0.230 & 0.220 & -- & -- & -- & -- \\
    NRMSE          & 0.152 & 0.143 & 0.171 & 0.177 & 0.177 & 0.147 & 0.159 & 0.138 & 2.07 & 1.27 & 0.958 & 1.25 \\
    $e_{99}$       & 0.701 & 0.667 & 0.796 & 0.748 & 0.742 & 0.621 & 0.676 & 0.585 & 5.31 & 4.99 & 4.14 & 4.86 \\
  \end{tabular}
  \caption{A priori errors of the surrogate for the catalytic configuration, on the held-out test set ($20\%$ of the samples, not used for training), for representative thermochemical outputs: mean relative error $\overline{e}$ (strictly positive outputs only), normalized root-mean-square error NRMSE, and 99th percentile $e_{99}$ of the normalized absolute error. All values are reported in $\%$.}
  \label{tab:apriori}
\end{table}

For this step, the local surrogate models described in Sect.~\ref{sec:RBF} are constructed for each of the $N_c$ communities identified, using $N_R = 500$ RBF centers per community, augmented with a ghost layer of $N_g = 1000$ nearest neighboring points from each of the adjacent communities.

Figure \ref{fig:modelerror} shows the relative error fields, $e_T = (\hat{T}-T)/T \, (\%)$, between the reference temperature field $T$ and the data-driven prediction $\hat{T}$, for the (a) non-catalytic and (b) catalytic-patch configurations. These fields are evaluated at all samples, from both the training $N_{\mathrm{train}}$ and test $N_{\mathrm{test}}$ sets; the errors on the two sets are almost identical (mean temperature error of $0.104\%$ and $0.109\%$, respectively, for the catalytic case), so the surrogate does not overfit the training data. In both configurations, the predicted temperature field agrees closely with the reference solution, with a mean relative error of about $0.1\%$ and local errors below about $2\%$. The catalytic-patch case exhibits slightly larger errors in the vicinity of the upstream patch boundary, where the rapid onset of surface catalysis gives rise to the strongest thermochemical gradients. Nevertheless, the relative error remains small, demonstrating that the proposed data-driven model accurately captures both the smooth and sharply varying thermochemical states.

Table~\ref{tab:apriori} extends this assessment to the other outputs of the thermochemical closure, for the catalytic configuration, which is the most demanding one, and evaluated on the test set only. For a given output $z$, with reference values $z_i$ and surrogate predictions $\hat z_i$ at the $N_\mathrm{test}$ samples of the test set, three error measures are reported: the mean relative error
$
  \overline{e} = {N_\mathrm{test}}^{-1} \sum_{i=1}^{N_\mathrm{test}}
                 {|\hat z_i - z_i|}/{|z_i|},
  \label{eq:emean}
$
the normalized root-mean-square error
$
  \mathrm{NRMSE} = {\Delta z}^{-1}
     \left( {N_\mathrm{test}}^{-1} \sum_{i=1}^{N_\mathrm{test}}
            \left(\hat z_i - z_i\right)^2 \right)^{1/2}
  \label{eq:nrmse}
$ and the 99th percentile $e_{99}$ of the normalized absolute errors $|\hat z_i - z_i|/\Delta z$, i.e.\ the value exceeded by only $1\%$ of the test samples, with $\Delta z = \max(z) - \min(z)$ denoting the range of the output over all samples. Since the production rates change sign, their relative error is ill-conditioned where they vanish, so $\overline{e}$ is reported only for strictly positive outputs. The two range-normalized measures remain well defined for all outputs. The thermodynamic and transport properties are predicted with mean relative errors below
$0.25\%$, for all species. The production rates are the most demanding outputs, with normalized root-mean-square errors of $1$--$2\%$, as commonly observed for reduced parameterizations of thermochemistry \citep{biglari2012}. Their largest errors ($e_{99}\sim 5\%$) are concentrated near the wall at the upstream edge of the catalytic patch, where the composition changes most rapidly, and affect only a small fraction of the samples. As shown in the next section, such error levels do not compromise the closed-loop prediction of the wall composition and temperature. For the non-catalytic configuration, the normalized errors of all outputs remain well below $1\%$.

\subsection{Model stability and performance}

\begin{figure}[!t]
\centering
\begin{tabular}{ll}
 (a) & (b) \\
\includegraphics[scale=0.43,trim=0 5 0 0,clip]{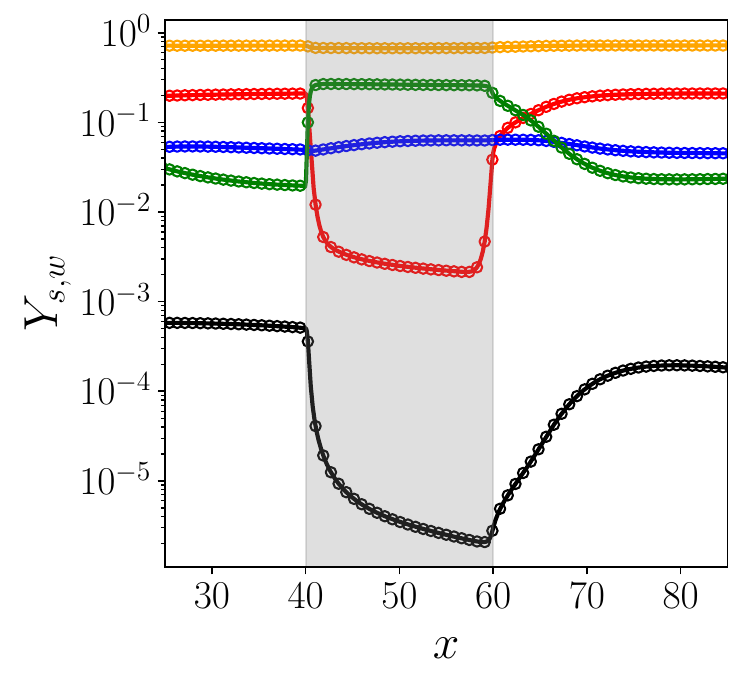} & 
\includegraphics[scale=0.43,trim=0 5 0 0,clip]{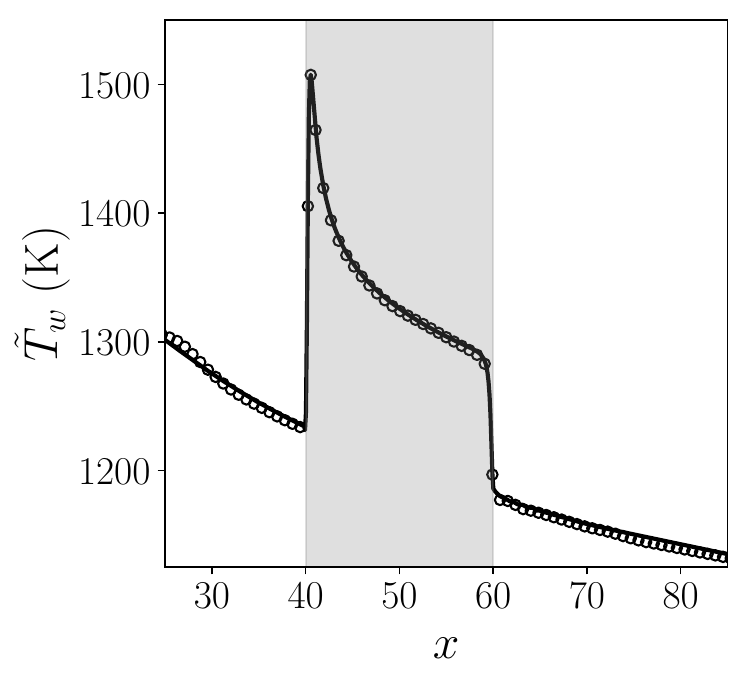} \\
 (c) & (d) \\
\includegraphics[scale=0.43,trim=0 0 0 8,clip]{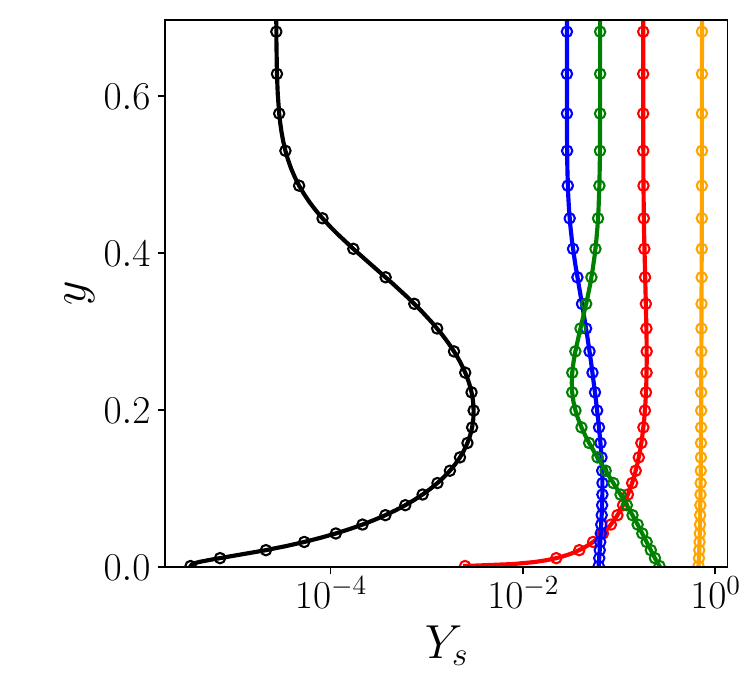} & 
\includegraphics[scale=0.43,trim=0 0 0 8,clip]{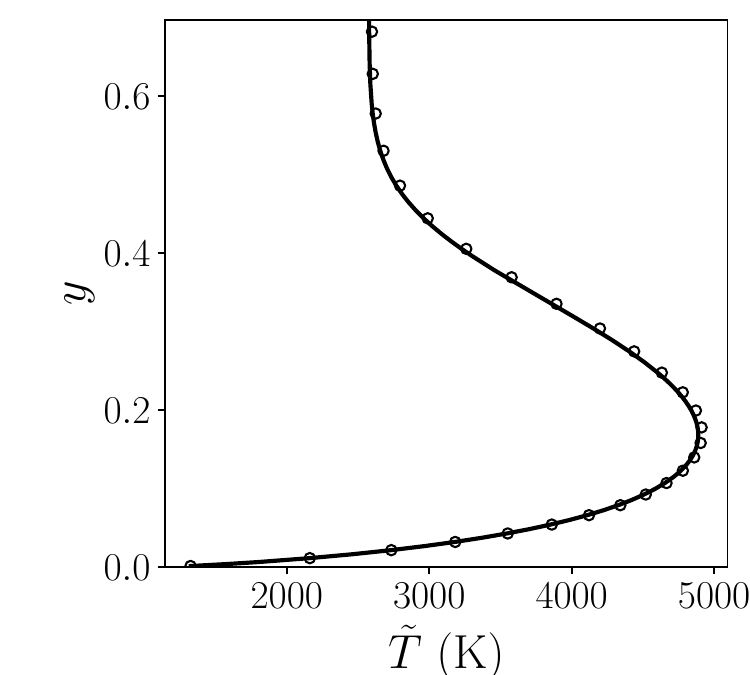} 
\end{tabular}
\caption{Closed-loop (\textit{a posteriori}) validation of the reduced-order model for the fully-catalytic case ($\Gamma_{\mathrm{O}} = 1$), obtained by substituting Mutation$^{++}$ Light for Mutation$^{++}$ throughout the $\textrm{MS}^{2}$ solver. Panels (a)-(d) and color coding are as in Fig.~\ref{fig:steadystate}. Solid lines denote the reference full-order Mutation$^{++}$ solution while open circles denote the corresponding Mutation$^{++}$ Light closed-loop prediction.}
\label{fig:stability}
\end{figure}

Having established the pointwise accuracy of the data-driven surrogate in an \textit{a priori} sense, we now assess its stability and performance when used to replace Mutation$^{++}$ entirely within a running simulation, i.e., in an \textit{a posteriori}, closed-loop setting. To this end, following \cite{scherding2023data}, every call to Mutation$^{++}$ within the flow solver is replaced by the corresponding Mutation$^{++}$ Light evaluation, without modifying the governing equations or the numerical solution procedure. The latter applies also to the surface balance Eqs.~\eqref{eqn:smb}-\eqref{eqn:seb}, which are solved at every stage exactly as in the full-order simulation. However, all the surface thermochemical properties required are extrapolated to the wall from Mutation$^{++}$ Light predictions, so that the surrogate closes the gas-phase equations as well as the surface balance (Appendix~\ref{app:catalyticBC}). 

The closed-loop simulation is initialized from the converged, steady-state solution obtained with the full-order model and is then advanced using Mutation$^{++}$ Light exclusively, for one flow-through time of the computational domain. We restrict this assessment to the catalytic case ($\Gamma_\mathrm{O} = 1$), as it constitutes the most demanding test of the reduced-order framework, combining the largest number of thermochemical communities and sharp wall-normal and streamwise gradients.

Figure \ref{fig:stability} compares the resulting closed-loop solution against the reference Mutation$^{++}$ solution of Fig.~\ref{fig:steadystate}, in terms of the wall species mass fractions and wall temperature, as well as profiles of species mass fractions and temperature extracted at $x = 50$. The data-driven simulation remains numerically stable throughout the simulation, with no evidence of drift, oscillation, or blow-up, despite the rapid variations in surface catalycity imposed at the patch
boundaries. The predicted fields overlay the reference solution closely across the entire domain, including in the near-wall region and at the upstream and downstream edges of the catalytic patch, where the thermochemical gradients are steepest. Note that this closed-loop test is in-sample. The surrogate is evaluated on the configuration from which its training data were sampled, and the simulation is initialised from the converged reference solution. It therefore assesses the consistency and closed-loop stability of the offline surrogate and does not assess predictive capability for unseen conditions, which requires the active-learning stage of RONAALP \citep{scherding2025}. 
Most importantly though for this study, this agreement shows that the surrogate can close the catalytic surface balance. The wall properties entering Eqs.~\eqref{eqn:smb}-\eqref{eqn:seb} are extrapolated from Mutation$^{++}$ Light predictions, the near-wall state that drives the balance is advanced with surrogate-closed fluxes, and the resulting wall state is fed back to the first cell (Appendix~\ref{app:catalyticBC}). Any surrogate error in the oxygen-depleted near-wall states created by catalysis (community $k=4$, Fig.~\ref{fig:clusters}) would thus appear directly in the predicted wall composition and temperature.

In terms of performance, the offline and online costs of the data-driven simulation are summarized in Table~\ref{tab:cost}, relative to the cost of the full-order simulation. Online, the cost of an iteration is dominated by the evaluation of the thermochemical closure, which accounts for about $90\%$ of it with Mutation$^{++}$. Replacing Mutation$^{++}$ with Mutation$^{++}$ Light reduces the cost of the closure by $55\%$, while the rest of the iteration is unchanged, so that the online runtime of the simulation is reduced by $50\%$. This profiling follows \citet{scherding2023data}, who showed that the evaluation cost of both libraries scales linearly with the number of thermochemical states, with a ratio of $0.45$ between the costs per state of the surrogate and of Mutation$^{++}$, and measured a $50\%$ reduction of the CPU time of their Mach-10 boundary-layer simulation, as well as a factor-of-two speed-up for their shock-wave/boundary-layer interaction. Both results, as in the present work, were obtained using Ramshaw's self-consistent effective binary diffusion model. Larger savings (up to $70\%$) were reported when Mutation$^{++}$ also solves the Stefan--Maxwell diffusion problem, which increases the share of the closure in the total cost. 

The offline cost consists of the extraction of the training states and of the construction of the surrogate. The training states are extracted by post-processing from a single snapshot of the converged solution, at negligible cost. The construction of the surrogate is dominated by the encoder training and the community detection, and costs less than $1\%$ of the full-order simulation; it is therefore recovered within $2\%$ of a data-driven simulation of the same length. 

\begin{table}[h]
  \centering
  \small
  \begin{tabular}{lcc}
    \hline
    \emph{Online} (per iteration) & M$^{++}$ & M$^{++}$ Light \\
    \hline
    Thermochemical closure & 0.90 & 0.41 \\
    Other operations (fluxes, time integration) & 0.10 & 0.10 \\
    Total & 1.00 & 0.51 \\
    \hline
    \emph{Offline} (relative to the full-order run) & \multicolumn{2}{c}{} \\
    \hline
    Extraction of the training states (one snapshot) & \multicolumn{2}{c}{negligible} \\
    Construction of the surrogate & \multicolumn{2}{c}{$< 1\%$} \\
    \quad of which encoder training & \multicolumn{2}{c}{$\approx 40\%$} \\
    \quad of which community detection & \multicolumn{2}{c}{$\approx 55\%$} \\
    \quad of which classifier and RBFs & \multicolumn{2}{c}{$\approx 5\%$} \\
    \hline
  \end{tabular}
  \caption{Relative costs of the data-driven simulation. Online: cost of one iteration of the flow solver, normalized by that with Mutation$^{++}$. Offline: cost of the construction of the surrogate relative to the full-order run, and breakdown of the construction time.}
  \label{tab:cost}
\end{table}

Taken together, these results demonstrate that the data-driven surrogate is accurate in an offline, pointwise sense, but also stable and computationally advantageous once embedded within the full nonlinear feedback loop of the CFD solver.

\section{Conclusions}\label{sec:conclusions}

In this paper, the RONAALP data-driven reduced-order framework has been extended to hypersonic reactive boundary-layer flows over surfaces exhibiting heterogeneous catalysis, a configuration not previously addressed within this modeling paradigm. The configuration under study consists of the post-shock boundary layer generated by a $M_{\infty} = 23.4$ hypersonic freestream flow over a $17^\circ$ wedge, yielding a boundary-layer edge Mach number of $M_e = 5.75$ and post-shock edge temperatures in excess of $3000\, \mathrm{K}$, sufficient to drive significant dissociation of molecular oxygen. The wall, maintained in radiative equilibrium at an emissivity of 0.9, is prescribed as non-catalytic everywhere except over a localized streamwise interval, where a catalytic patch is imposed, with a rapid but continuous transition in recombination coefficient at the patch boundaries, as required by the high-order discretisation of the solver. This configuration, representative of the catalytic transitions encountered at material junctions in real thermal protection systems, was solved at its steady, laminar state using the full compressible multi-species Navier–Stokes equations coupled to Mutation++ for thermochemical closure.

The full-order simulations reveal the characteristic physical signature of localized surface catalysis. Within the catalytic patch, the imposed recombination reaction strongly depletes atomic oxygen at the wall while enriching molecular oxygen, and, through multicomponent diffusion and the mass-fraction constraint, indirectly redistributes atomic and molecular nitrogen and nitric oxide. Due to the exothermic nature of catalytic reactions, these compositional changes are accompanied by a rise in wall temperature of approximately $100 \, \mathrm{K}$ across the patch, with a pronounced overshoot in wall temperature occurring at the upstream patch boundary, as excess atomic oxygen convected from upstream non-catalytic conditions is suddenly exposed to an active catalytic surface. Those effects remain confined to the near-wall region of the boundary layer and persist downstream of the patch over a convective-diffusive relaxation length, leaving the outer boundary layer and freestream composition largely unaffected. These findings confirm that localized changes in surface catalycity generate sharp, spatially localized gradients in diffusion fluxes and surface heat transfer, motivating the investigation of whether the data-driven reduction framework can adapt to such variations and accurately identify the physically relevant flow regions.

To capture this behavior within the reduced-order model, the RONAALP framework required one methodological extension. The nonlinear dimensionality reduction step was augmented with a Sammon-type stress penalty that explicitly preserves the pairwise-distance structure of the input space in the latent embedding. This addition proved essential, since the enlarged thermochemical state space induced by catalytic wall reactions was shown to promote severe topological folding of the latent manifold when left unregularized, degrading the physical consistency of the subsequent clustering and surrogate stages. 

The full RONAALP pipeline, i.e. dimensionality reduction, community clustering, and local surrogate modeling, was applied to both the non-catalytic and catalytic-patch datasets. Notably, the community-detection stage identified an additional thermochemical regime induced by the catalytic reactions: while the non-catalytic wall yields three clusters associated with the near-wall, high-temperature, and freestream regions, the catalytic patch gives rise to a fourth, distinct community, despite comparable temperature levels to the surrounding non-catalytic wall. This result confirms that the latent representation already encodes the distinct thermochemical regime induced by surface recombination rather than temperature alone.

The resulting reduced-order model reproduces the reference Mutation++ solution with high fidelity in an \textit{a priori}, pointwise sense, with mean relative errors below $0.25\%$ for the thermodynamic and transport properties and normalized errors of a few percent for the production rates, including the vicinity of the catalytic patch boundaries where thermochemical gradients are most severe. Beyond this pointwise accuracy, substituting the trained surrogate for Mutation++ throughout a full closed-loop simulation was shown to remain numerically stable, with no evidence of drift or blow-up, while reducing the online simulation cost by $50\%$ relative to the full thermochemical library. Together, these a priori and a posteriori results show that the offline RONAALP surrogate, trained on Mutation$^{++}$ data, can represent and stably close a hypersonic boundary layer with catalytic wall boundary conditions. It captures the sharp gradients in wall species mass fractions, diffusion fluxes and surface heat transfer at the catalytic transition, at half the online computational cost of the full library.

These findings extend the applicability of the RONAALP framework from the chemically inert, adiabatic-wall configurations previously considered to flows over catalytic walls, featuring localized catalytic transitions representative of realistic thermal
protection system designs. The present assessment is in-sample, and it is not a demonstration of predictive capability for unseen conditions. As established by \citet{scherding2025}, out-of-sample use of the surrogate requires its online active-learning stage. The present results show that catalytic boundary conditions create a new thermochemical regime, identified as a separate community, rather than only new states within existing regimes. The online stage, which currently keeps the encoder and the community partition fixed, must therefore be extended to create new communities, and possibly to adapt the latent manifold, during the simulation. Future work will develop this extension and apply it to out-of-sample catalytic configurations (different recombination coefficients and patch geometries) and to unsteady flows over catalytic surfaces, where instabilities and transition may drive the flow through thermochemical states not seen during training.

\backmatter

\bmhead{Funding} 
This work is part of the PostGenAI@Paris project and was supported by the French government through the Agence Nationale de la Recherche (ANR) under the France 2030 program (reference: ANR-23-IACL-0007), as well as by the European Space Agency (ESA) through the Open Space Innovation Platform (OSIP).

\bmhead{Author contributions}
K.S.: conceptualization, methodology, simulations, analysis, data curation, visualization, validation, writing of original draft and editing 
L.W.: supervision, validation, review. 
T.M.: supervision, conceptualization, methodology, validation, review. 
P.S.: supervision, conceptualization, methodology, validation, review. 
T.S.: supervision, conceptualization, methodology, validation, review and editing.

\bmhead{Data availability}
The data used to train the reduced-order model are available upon request.

\bmhead{Acknowledgements}
The authors would like to thank Dr. Michele Capriati (VKI) for fruitful discussions, valuable assistance, and for providing relevant materials that contributed to this work. 

\section*{Declarations}
\bmhead{Competing interests}
The authors declare no competing interests.




\begin{appendices}

\section{Post-shock boundary-layer edge conditions}
\label{secA:edge}

This appendix describes how the boundary-layer edge state (Table~\ref{tab:freestream_caseA_essential}) is derived from the freestream conditions, including the thermochemical assumptions and the initialization of the edge composition. The procedure consists of two steps. First, the shock angle, edge Mach number, pressure and temperature are determined using oblique-shock relations for a calorically perfect gas. Second, the resulting mechanical state is supplemented with a thermochemical state at chemical equilibrium, computed using Mutation$^{++}$.

The shock is computed assuming a calorically perfect gas with a constant ratio of specific heats, $\gamma = 1.4$. This assumption corresponds to frozen chemical composition and neglects vibrational excitation across the shock. For a wedge with a deflection angle $\theta$, the shock angle $\beta$ is determined from the $\theta$--$\beta$--$M$ relation \citep{andersonbook},
\begin{equation}
  \tan\theta
  = 2\cot\beta\,
    \frac{M_\infty^2\sin^2\beta - 1}
         {M_\infty^2\left(\gamma + \cos 2\beta\right) + 2},
  \label{eqA:tbm}
\end{equation}
of which the weak-shock solution is retained. The normal-shock relations are
then applied to the freestream Mach number normal to the shock,
$
  M_{n,\infty} = M_\infty \sin\beta,
  \label{eqA:Mn1}
$
which gives the following pressure, density and temperature ratios
\begin{equation}
  \frac{\tilde p_e}{\tilde p_\infty}
    = 1 + \frac{2\gamma}{\gamma+1}\left(M_{n,\infty}^2 - 1\right),
  \quad
  \frac{\tilde\rho_e}{\tilde\rho_\infty}
    = \frac{(\gamma+1)\,M_{n,\infty}^2}{(\gamma-1)\,M_{n,\infty}^2 + 2},
  \quad
  \frac{\tilde T_e}{\tilde T_\infty}
    = \frac{\tilde p_e}{\tilde p_\infty}\,
      \frac{\tilde\rho_\infty}{\tilde\rho_e},
  \label{eqA:ratios}
\end{equation}
and the post-shock Mach number
\begin{equation}
  M_{n,e}^2 = \frac{(\gamma-1)\,M_{n,\infty}^2 + 2}
                   {2\gamma\,M_{n,\infty}^2 - (\gamma-1)},
  \qquad
  M_e = \frac{M_{n,e}}{\sin(\beta-\theta)} .
  \label{eqA:Me}
\end{equation}
For $M_\infty = 23.4$ and $\theta = 17^\circ$,
Eqs.~\eqref{eqA:tbm}--\eqref{eqA:Me} give the following post-shock values and edge-to-freestream ratios
\begin{equation}
  \beta = 20.90^\circ,\quad
  M_{n,\infty} = 8.35,\quad
  \frac{\tilde p_e}{\tilde p_\infty} = 81.1,\quad
  \frac{\tilde\rho_e}{\tilde\rho_\infty} = 5.60,\quad
  \frac{\tilde T_e}{\tilde T_\infty} = 14.5,\quad
  M_e = 5.75 .
  \label{eqA:shockvalues}
\end{equation}
Given the freestream state of Table~\ref{tab:freestream_caseA_essential}, these relations yield $\tilde p_e \simeq 1782$~Pa and
$\tilde T_e \simeq 3106$~K.

The edge composition is set to the local chemical-equilibrium composition at $(\tilde T_e, \tilde p_e)$, computed with Mutation$^{++}$ for the five-species air mixture, or
\begin{equation}
  Y_{s,e} = Y_s^{\mathrm{eq}}\!\left(\tilde T_e, \tilde p_e\right),
  \qquad s \in \mathcal{S}.
  \label{eqA:Yeq}
\end{equation}
All remaining edge properties are then evaluated with Mutation$^{++}$ at this state, $(\tilde T_e, \tilde p_e, Y_{s,e})$. These include the density $\tilde\rho_e$, the frozen isentropic exponent $\gamma_e$, the frozen speed of sound $\tilde a_e$, the viscosity $\tilde\mu_e$ and the edge velocity $\tilde U_e = M_e \tilde a_e$. These reference quantities are listed in Table~\ref{tabA:edge-appendix}. In particular, the edge density follows from the equation of state of the partially dissociated mixture, $\tilde\rho_e = \tilde p_e/(\tilde R_e \tilde T_e)$, with $\tilde R_e = \mathcal{R}/\mathcal{M}_e = 333.8$~J/(kg\,K). Since this gas constant is higher than that of undissociated air, $\tilde\rho_e$ is about $14\%$ lower than the calorically perfect value of Eq.~\eqref{eqA:rhoratio} ($5.60\,\tilde\rho_\infty \simeq 2.0\times10^{-3}$~kg/m$^3$). Likewise, $\tilde U_e$ is about $6\%$ higher than the calorically perfect value ($\simeq 6430$~m/s), owing to the higher speed of sound of the dissociated mixture.

\begin{table}[h]
  \centering
  \begin{tabular}{cccccc}
    \hline
    $Y_{\mathrm{N}}$ & $Y_{\mathrm{O}}$ & $Y_{\mathrm{NO}}$ &
    $Y_{\mathrm{N_2}}$ & $Y_{\mathrm{O_2}}$ & \\
    $9.3\times10^{-5}$ & $0.179$ & $0.029$ & $0.729$ & $0.063$ & \\
    \hline
    $\tilde\rho_e$ (kg/m$^3$) & $\mathcal{M}_e$ (kg/mol) & $\gamma_e$ &
    $\tilde a_e$ (m/s) & $\tilde\mu_e$ (Pa\,s) & $\tilde U_e$ (m/s) \\
    $1.72\times10^{-3}$ & $0.0249$ & $1.345$ & $1181$ &
    $1.02\times10^{-4}$ & $6791$ \\
    \hline
  \end{tabular}
  \caption{Thermochemical edge state used in the simulations, evaluated with Mutation$^{++}$ (five-species air, chemical equilibrium) at $\tilde T_e = 3106$~K and $\tilde p_e = 1782$~Pa.}
\label{tabA:edge-appendix}
\end{table}

At the edge, the gas is thus partially dissociated: about 70\% of the oxygen mass is in atomic form, while nitrogen remains essentially molecular. The flow field is initialised with the self-similar boundary-layer solution and the uniform composition $Y_{s,e}$. The same composition defines the reference state towards which the inflow and outer sponge zones relax. Since $Y_{s,e}$ is an equilibrium composition, $\dot\omega_s = 0$ at the edge state. The outer flow therefore remains chemically steady and composition changes are confined to the boundary layer.

\section{Thermochemical and kinetic models}\label{appendix:thermochemical-kinetics}
This appendix details the thermochemical kinetics model used to compute the gas-phase chemical source terms. The formulation is based on reversible finite-rate reactions and Arrhenius-type reaction rates. Each reaction $r$ in the gas-phase is written as
\begin{equation}
\sum_{s \in \mathcal{S}}  \nu'_{r,s} s
\;\overset{k_{f,r}}{\underset{k_{b,r}}{\rightleftharpoons}}\;
\sum_{s \in \mathcal{S}} \nu''_{r,s} s ,
\quad \forall r \in \mathcal{R}, 
\end{equation}
where $\nu'_{r,s}$ and $\nu''_{r,s}$ are the stoichiometric coefficients for reactants and products in
reaction $r$ for species $s$ and is characterized by the forward rate $k_{f,r}$ and the backward rate $k_{b,r}$. These, in turn, are obtained
according to experimentally or theoretically calibrated Arrhenius formulas \citep{mutation1}. The net mass production rates $\dot{\omega}_s$ are governed by the law of mass action
\begin{equation}
\dot{\omega}_s = M_s \sum_{r \in \mathcal{R}} 
\left( \nu''_{r,s} - \nu'_{r,s} \right)
\left[
k_{f,r} \prod_{j \in \mathcal{S}} \left( \frac{\rho_j}{M_j} \right)^{\nu'_{r,j}}
-
k_{b,r} \prod_{j \in \mathcal{S}} \left( \frac{\rho_j}{M_j} \right)^{\nu''_{r,j}}
\right]
[\text{M}]_r 
\end{equation}
where $[\text{M}]_r=\sum_{j \in \mathcal{S}} Z_{r,j} \frac{\rho_j}{M_j}$ represents the effective third-body concentration, with $Z_{r,j}$ being the third-body efficiency of species $j$ in reaction $r$. Note that for the elementary reactions that do not involve a third body, $[\text{M}]_r = 1$ by definition.

Local species concentration gradients drive diffusion, represented in the governing species Eq.~\eqref{eqn:species} by the diffusive flux $\mathbf{J}_s$. Multicomponent diffusion is modeled in this paper using the Self-Consistent Effective Binary Diffusion model \citep{ramshaw}, a formulation based on Fick's law, corrected so as to enforce overall mass conservation of the diffusive fluxes by solving the following coupled system of equations
\begin{equation}\label{eqn:ficklaw}
\mathbf{J}_s = -c M_s D_s \nabla Y_s + c Y_s \sum_{k \in \mathcal{S}} M_k D_k \nabla Y_k
\end{equation}
where $c = \sum_{s \in \mathcal{S}} \rho_s / M_s$ denotes the total molar concentration and
\begin{equation}
D_s =
\frac{
\sum_{r \ne s} \left( Y_r / M_r \right)
}{
\sum_{r \ne s} \left( Y_r / (M_r D_{s,r}) \right)
}
\end{equation}
is the effective diffusion coefficient, with $D_{s,r}$ denoting the binary diffusion coefficients obtained using Wilke’s mixing rule.

At all times, the following constraints are satisfied for the kinetic and diffusive terms in the gas phase
\begin{equation}
\sum_{s \in \mathcal{S}} \dot{\omega}_s = 0,
\qquad
\sum_{s \in \mathcal{S}}  \mathbf{J}_s = 0.
\end{equation}

\section{Implementation and verification of catalytic boundary conditions}\label{app:catalyticBC}

This appendix describes how the catalytic boundary conditions (Eqs.~\eqref{eqn:smb}--\eqref{eqn:seb}) are discretized and enforced in the MS$^2$ solver but also verifies their implementation against an analytical solution. At each wall point, these equations form a nonlinear algebraic system for the wall composition and temperature, which is
solved with Newton's method at every Runge--Kutta stage, in both the full- and reduced-order simulations.

Since the governing equations are discretized on a staggered grid (Sect.~\ref{sec:numerical}), conservative variables are stored at the cell centres and no variable is stored at the wall itself. Wall values are therefore obtained by a one-sided extrapolation from the first four cell centres, and, conversely, a prescribed wall value $\phi_w^{\star}$ is enforced by modifying the first cell only,
\begin{equation}
  \phi_w=\sum_{j=1}^{4}a_j\,\phi_j,
  \qquad
  \phi_1=\frac{1}{a_1}\Bigl(\phi_w^{\star}-\sum_{j=2}^{4}a_j\,\phi_j\Bigr),
  \label{eqA:extrap}
\end{equation}
where the coefficients $a_j$ are those of the fourth-order boundary closure of the compact scheme \citep{nagarajan2}. The same pair of operations is used for all wall boundary conditions of the solver.

At each wall point, the vector of unknowns of the surface balance consists of the wall mass fractions and the wall temperature,
$\boldsymbol{\chi} =(Y_{s,w},T_w)^{\top}, s\in\mathcal{S}$, the wall pressure $p_w$ being held fixed at its extrapolated value, so that the partial densities follow from the equation of state as $\rho_{s,w}=Y_{s,w}\,\rho_w$, with $\rho_w=p_w/\bigl(\mathcal{R}_u T_w\sum_{s}Y_{s,w}/M_s\bigr)$. The wall-normal gradients are approximated between the wall and the first cell centre, located at a distance $\Delta y_1$ from the wall, as $\partial_n Y_s|_w\simeq (Y_{s,1}-Y_{s,w})/\Delta y_1$ and $\partial_n T|_w\simeq (T_1-T_w)/\Delta y_1$, where $T_1$ is a closure output at the first cell. The thermochemical properties required at the wall, $(D_s,\kappa_w,h_s)$, are not evaluated by additional library calls. Instead, they are extrapolated with Eq.~\eqref{eqA:extrap} from their values at the cell centres, provided by Mutation$^{++}$ in the full-order simulations and by Mutation$^{++}$ Light in the reduced-order ones, and are held fixed during the Newton iterations.

The wall diffusion fluxes follow the same effective binary diffusion model as the gas phase (Eq.~\eqref{eqn:ficklaw}) evaluated at the wall,
\begin{equation}
  \mathbf{J}_{s,w}\!\cdot\boldsymbol{\hat{n}}_w
  = -\,c\,M_s D_s \left.\frac{\partial Y_s}{\partial n}\right|_w
    + c\,Y_{s,w}\sum_{k\in\mathcal{S}} M_k D_k \left.\frac{\partial Y_k}{\partial n}\right|_w ,
  \label{eqA:Jw}
\end{equation}
where $c$ follows from the current iterate $\boldsymbol{\chi}$, while the catalytic source terms $\dot{\omega}_{w,s}$ of Eq.~\eqref{eqn:wallsource} are explicit functions of $(\rho_{\mathrm{O},w},T_w)$ and of the local $\gamma_\mathrm{O}(x)$, and require no closure. Equations~\eqref{eqn:smb}--\eqref{eqn:seb} are thus recast as $\mathbf{F}(\boldsymbol{\chi})=(F_s(\boldsymbol{\chi}),F_E(\boldsymbol{\chi}))^\top=\mathbf{0}$, with
\begin{equation}
  \begin{aligned}
  F_s &= \mathbf{J}_{s,w}\!\cdot\boldsymbol{\hat{n}}_w-\dot{\omega}_{w,s},
  \qquad s\in\mathcal{S},\\
 F_E &= \kappa_w\left.\frac{\partial T}{\partial n}\right|_w
     - \sum_{s\in\mathcal{S}} h_{s,w}\,\mathbf{J}_{s,w}\!\cdot\boldsymbol{\hat{n}}_w
     - \sigma\epsilon T_w^{4},
  \end{aligned}
  \label{eqA:F}
\end{equation}
written here in dimensional form, in which the system is solved. Newton's method updates the iterate as
$\boldsymbol{\chi}^{(k+1)}=\boldsymbol{\chi}^{(k)}-\mathsf{J}^{-1}\mathbf{F}(\boldsymbol{\chi}^{(k)})$, where the Jacobian $\mathsf{J}=\partial\mathbf{F}/\partial\boldsymbol{\chi}$ is approximated by forward finite differences. Since the diffusion fluxes and the catalytic source terms both sum to zero over the species, the species block of $\mathsf{J}$ is singular; as is standard, it is regularized by the rank-one term $a\,\mathbf{1}\mathbf{1}^{\top}$, with $a$ the largest diagonal entry of the block, which also enforces $\sum_s Y_{s,w}=1$. The iterations start from the extrapolated wall composition and the wall temperature of the previous stage, and are stopped when $\|\Delta\boldsymbol{\chi}\|_\infty<10^{-12}$.

Upon convergence, the catalytic wall state $(\rho_{s,w}^{\star},T_w^{\star})$ is imposed through the second relation of Eq.~\eqref{eqA:extrap}, by modifying the transported partial densities, the mixture density $\rho_w^{\star}=\sum_s\rho_{s,w}^{\star}$, the momentum (no-slip condition) and the energy of the first cell; since the wall velocity vanishes, the latter reduces to the internal energy $(\rho e)_w^{\star}=\sum_s\rho_{s,w}^{\star}\,e_s(T_w^{\star})$. The thermochemical closure $\hat{\mathbf{z}}=(\mu,\kappa,p,T,D_s,h_s,\dot{\omega}_s)$ is then evaluated at all cell centres, including the modified first cell, with Mutation$^{++}$ in the full-order simulations or Mutation$^{++}$ Light in the reduced-order ones, and the solution is advanced to the next stage.

The implementation is verified by solving the diffusive transport of a binary mixture in a one dimensional tube, with catalytic conditions applied at one end of the domain. Following the analytical test problem of \citet{baskaya}, nitrogen recombination is considered for this verification, while the same catalytic boundary-condition equations apply to the oxygen recombination considered in this study. The setup of the test case is summarized in Table \ref{tab:1dtube} and consists of a binary nitrogen mixture $\mathcal{S}=\{ \mathrm{N},\mathrm{N}_2 \}$ along a tube of length $\tilde{L}=0.2~m$. The transport along the tube is driven solely by molecular diffusion, with no advective contribution ($\tilde{u}=0\, m/s$). The left end of the tube ($\tilde{x}=0~m$) at pressure $\tilde{p}_\infty=100~\mathrm{Pa}$ and temperature $\tilde{T}=3000~\mathrm{K}$ corresponds to a reservoir of nitrogen particles, with the mass fractions satisfying $Y_\mathrm{N}=1$ and $Y_\mathrm{N_2}=0$. On the right side ($\tilde{x}=0.2~m$) a catalytic boundary condition is imposed, with the catalytic wall ($w$) promoting the recombination of free nitrogen atoms, i.e. $\mathrm{N}+\mathrm{N} + w \to \mathrm{N}_2 +w $. The rate of chemical reactions is controlled by the recombination coefficient $\gamma_\mathrm{N}$, which in this test study takes values $\gamma_{\mathrm{N}} = \{ 0.01,0.1,1\}$ and ranges from partially to fully catalytic conditions. Note that there is no pressure or temperature gradient along the tube ($\tilde{p}_w/\tilde{p}_\infty = \tilde{T}_w/\tilde{T}_\infty = 1$).

\begin{figure}[!t]
    \centering
    \includegraphics[width=0.4\textwidth]{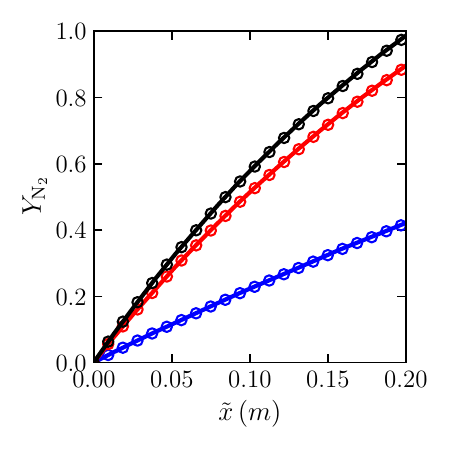} 
    \vspace{-1em}
    \caption{Mass fraction of molecular nitrogen $Y_{\mathrm{N_2}}$ as a function of the dimensional distance $\tilde{x}$, obtained using the MS$^2$ solver (open circles) and compared with the analytical results (solid lines) for $\gamma_\mathrm{N} = 0.01$ (blue), $\gamma_\mathrm{N} = 0.1$ (red), and $\gamma_\mathrm{N} = 1$ (black).}
    \label{fig:1dtube}
\end{figure}

\begin{table}[!t]
\centering
\small
\setlength{\tabcolsep}{2pt}
\begin{tabular}{c c c c c c c c c}
$\tilde{L}\,(\mathrm{m})$ &
$\tilde{u}\,(\mathrm{m/s})$ &
$\tilde{p}_\infty\,(\mathrm{Pa})$ &
$\tilde{T}_\infty\,(\mathrm{K})$ &
$\tilde{p}_w/\tilde{p}_\infty$ &
$\tilde{T}_w/\tilde{T}_\infty$ &
$Y_{\mathrm{N}_2}(\tilde{x}=0\,\mathrm{m})$ &
$Y_{\mathrm{N}}(\tilde{x}=0\,\mathrm{m})$ &
$\gamma_{\mathrm{N}}$ \\
\hline
$0.2$ & $0$ & $100$ & $3000$ & $1$ & $1$ &
$0$ & $1$ & $\{0.01,\,0.1,\,1\}$
\end{tabular}
\caption{Setup parameters and boundary conditions for the 1-D catalytic diffusion problem.}
\label{tab:1dtube}
\end{table}

Figure \ref{fig:1dtube} shows the mass fractions of molecular nitrogen $Y_{\mathrm{N}_2}$ as a function of the dimensional distance from the reservoir $\tilde{x}$, for the three different recombination coefficients ($\gamma_\mathrm{N}=1$ in black, $\gamma_\mathrm{N}=0.1$ in red and $\gamma_\mathrm{N}=0.01$ in blue). Since the mixture is binary, diffusion is modeled using Fick’s law, consistent with the analytical solution reported by \citet{baskaya}. Solid lines represent the analytical solutions while open circles indicate the steady-state results obtained with the MS$^2$ solver, after convergence to a steady state. As expected, increasing the recombination coefficient $\gamma_{\mathrm{N}}$ leads to higher values of $Y_{\mathrm{N_2}}$ at the wall, which reaches unity in the fully catalytic case. The agreement between the numerical and analytical solutions verifies the implementation.

\section{Effect of the stress penalty on manifold topology }\label{app:topology}

\begin{figure}[!t]
\begin{tabular}{ll}
\includegraphics[scale=0.6,trim=45 15 70 60,clip]{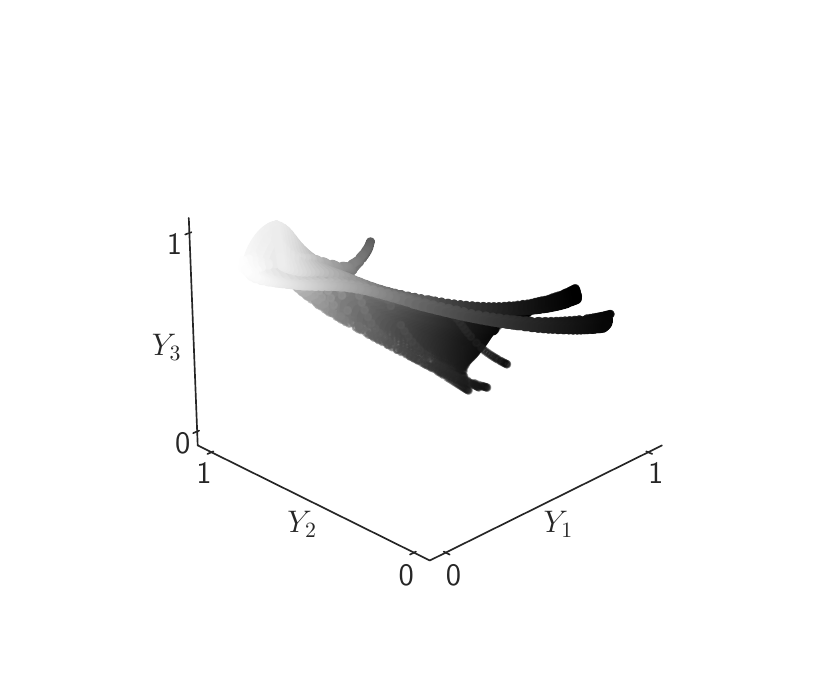} &
\raisebox{0.7em}{\includegraphics[scale=0.6,trim=0 10 0 10,clip]{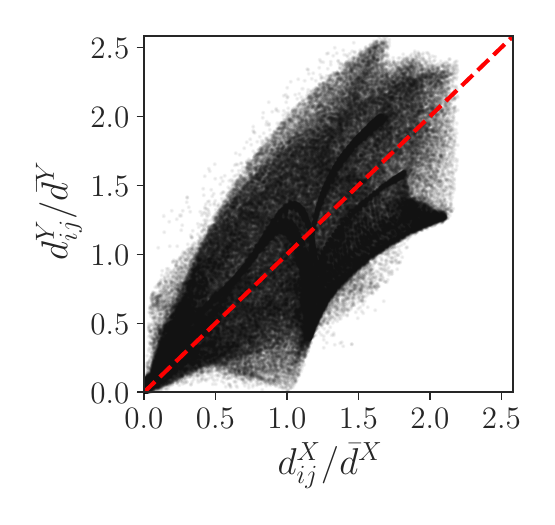}} \\[-1em]
\includegraphics[scale=0.6,trim=45 15 70 60,clip]{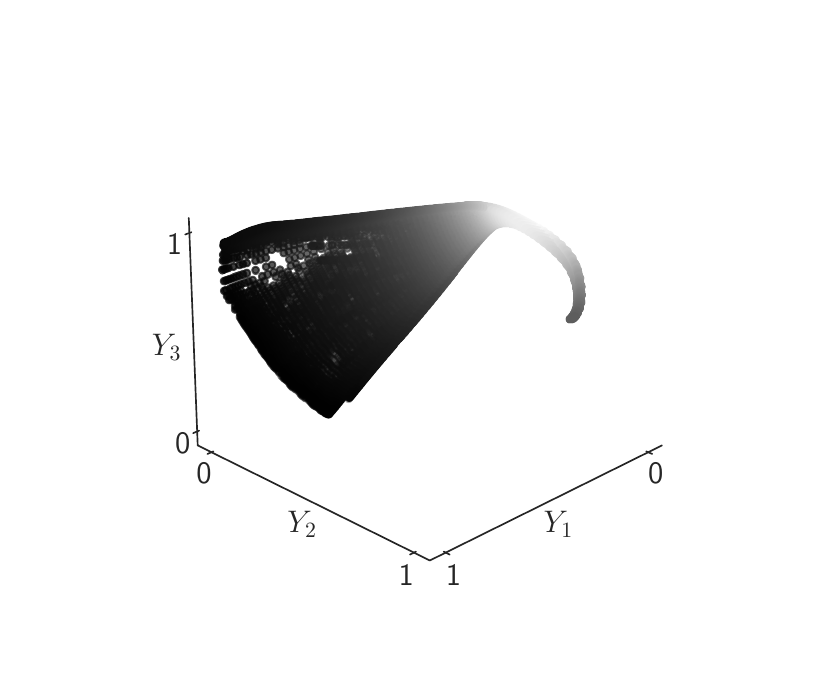} &
\raisebox{0.7em}{\includegraphics[scale=0.6,trim=0 10 0 10,clip]{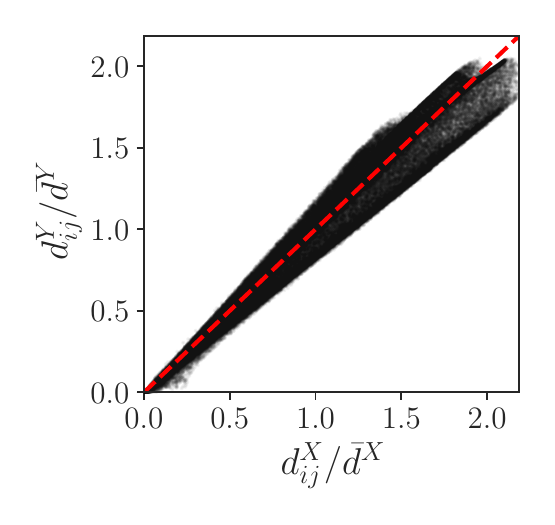}} \\[-1em]
\includegraphics[scale=0.6,trim=45 15 70 60,clip]{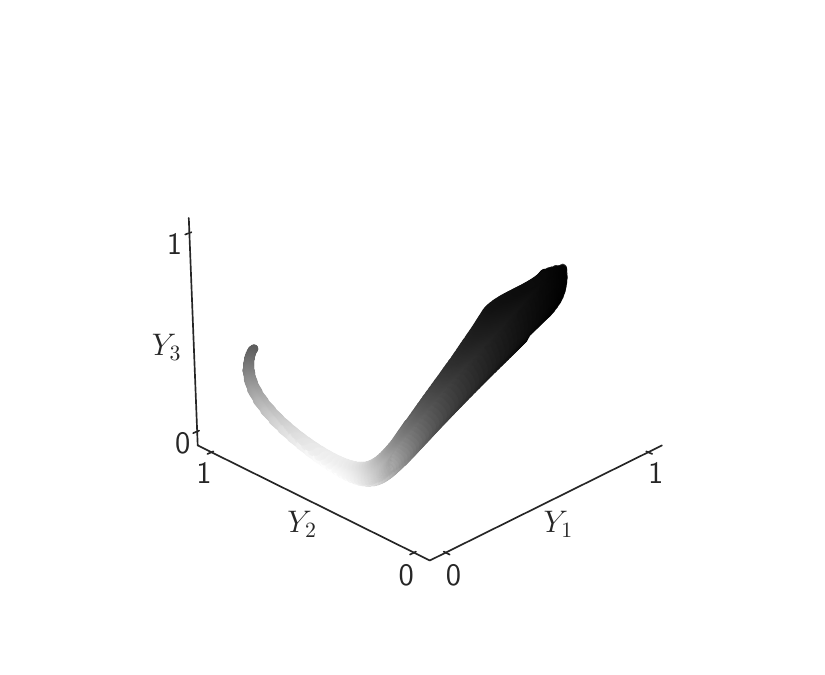} &
\raisebox{0.7em}
{\includegraphics[scale=0.6,trim=0 10 0 10,clip]{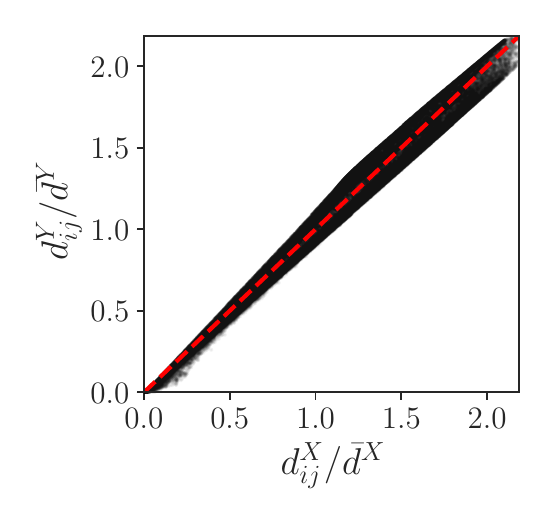}}
\end{tabular}
\caption{Effect of $\lambda_{\mathrm{stress}}$ on the manifold topology. (Left) manifolds in latent space ($Y_1,Y_2,Y_3$) and (right) Shepard diagrams of normalized pairwise distances in the input
space ($d^{X}_{ij}/\bar{d}^{X}$) versus the latent space ($d^{Y}_{ij}/\bar{d}^{Y}$), obtained for (from top to bottom) $\lambda_\mathrm{stress}=0$, $\lambda_\mathrm{stress}=1$ and $\lambda_\mathrm{stress}=10^3$. The red dashed line in the Shepard diagrams marks perfect distance preservation. Present results correspond to the catalytic dataset.}
\label{fig:lambdastress}
\end{figure}

In this section, we examine how the inclusion of the Sammon-type stress penalty defined in Eq.~\eqref{eq:ioe_loss} influences the topological structure of the resulting latent representation. Focusing only on the catalytic dataset, Fig.~\ref{fig:lambdastress} illustrates the effect of the stress weight $\lambda_{\mathrm{stress}}$ on the topology of the latent embedding, comparing three values, namely $\lambda_\mathrm{stress}=0$, $\lambda_\mathrm{stress}=1$ and $\lambda_\mathrm{stress}=10^3$, spanning the unregularized case to a strongly regularized one. For each value, panels on the left show the resulting manifold in $(Y_1,Y_2,Y_3)$ while panels on the right show the corresponding Shepard diagram of normalized pairwise distances, $d^{X}_{ij}/\bar{d}^{X}$ versus $d^{Y}_{ij}/\bar{d}^{Y}$, with the red dashed line marking perfect distance preservation.

For $\lambda_{\mathrm{stress}}=0$ (Fig.~\ref{fig:lambdastress}, top row), the encoder is trained with no distance-preservation penalty and is free to arrange $\mathbf{Y}$ solely to minimize $\mathcal{L}_{\mathrm{recon}}$. The resulting manifold exhibits strong folding, with self-intersections and disconnected regions of the input space overlapping on the same regions of the latent space. This is consistent with the corresponding Shepard diagram, in which a substantial fraction of points lie well away from the diagonal, a characteristic signature of topological folding. For $\lambda_{\mathrm{stress}}=1$ (Fig.~\ref{fig:lambdastress}, middle row), corresponding to equal weighting of reconstruction accuracy and topology preservation, the manifold unfolds into a single, smooth, self-consistent sheet. The corresponding Shepard diagram tightens considerably around the diagonal, indicating that the pairwise-distance structure of the input space is now largely preserved in the latent embedding. Increasing the weight to large values, such as $\lambda_{\mathrm{stress}}=10^3$ (Fig.~\ref{fig:lambdastress}, bottom row), drives the embedding toward a near-isometric copy of the input space, as evidenced by the Shepard diagram approaching a complete collapse onto the diagonal. However, this comes at the cost of restricting the encoder's freedom to nonlinearly reorganize the latent coordinates for reconstruction accuracy. 
\begin{table}[h]
  \centering
  \begin{tabular}{lccccc}
    $\lambda_\mathrm{stress}$ & mean $|e_T|$ (\%) & max $|e_T|$ (\%) & $\mathrm{NRMSE}_Z$ & $\rho_S$ & ARI \\
        \hline
       0   & $0.076 \pm 0.006$ & $1.28 \pm 0.23$ & $0.020 \pm 0.002$ & $0.69 \pm 0.10$ & $0.49 \pm 0.09$ \\
       0.1 & $0.104 \pm 0.011$ & $5.64 \pm 3.30$ & $0.030 \pm 0.002$ & $0.98 \pm 0.01$ & $0.94 \pm 0.02$ \\
       1   & $0.109 \pm 0.008$ & $2.36 \pm 0.80$ & $0.031 \pm 0.002$ & $0.96 \pm 0.04$ & $0.95 \pm 0.01$ \\
       10  & $0.174 \pm 0.007$ & $7.15 \pm 0.27$ & $0.056 \pm 0.002$ & $0.96 \pm 0.04$ & $0.77 \pm 0.13$ \\
       100 & $0.208 \pm 0.011$ & $6.39 \pm 0.32$ & $0.066 \pm 0.001$ & $0.97 \pm 0.02$ & $0.98 \pm 0.01$ \\
  \end{tabular}
    \caption{Effect of the stress weight $\lambda_\mathrm{stress}$ ($d=3$) on the catalytic dataset. Mean  and maximum relative temperature error $|e_T|$ (\%) and normalized root-mean-square error of the outputs $\mathrm{NRMSE}_Z$ on the test set, Spearman correlation $\rho_S$ between pairwise input and latent distances, and adjusted Rand index (ARI) between the community partitions obtained with different initializations. Mean $\pm$ standard deviation over three trainings.}
    \label{tabB:lambda}
\end{table}

To quantify these observations, the complete reduced-order model was trained
for $\lambda_\mathrm{stress} = 0, 0.1, 1, 10$ and $100$, for the catalytic dataset and with all other parameters unchanged, three times for each value with different random initializations. Table~\ref{tabB:lambda} reports the
mean and maximum relative temperature error, the normalized root-mean-square
error of the outputs, the Spearman correlation $\rho_S$ between the pairwise
distances in the input and latent spaces, and the adjusted Rand index (ARI)
between the community partitions obtained with different initializations.
The penalty does not improve the pointwise accuracy: for the catalytic
dataset, the unregularized model is slightly more accurate than the model
with $\lambda_\mathrm{stress}=1$, the maximum error being generally located near the patch boundaries in both cases. Its effect is on the structure of the latent space: $\rho_S$ increases from $0.69$ to $0.96$ and the community
partition becomes reproducible, with an ARI increasing from $0.49$ to $0.95$. Without the penalty, the folded manifold is arranged differently by each training, so that the communities, and hence their physical interpretation, depend on the initialization. For $\lambda_\mathrm{stress} \ge 10$, the errors increase, by a factor of about three for the maximum of the catalytic temperature error. The community-classification error of the random forest
remains below $0.3\%$ for all values, and the penalty increases the training
time by about $25\%$ while leaving the online cost of the surrogate unchanged.

\section{Choice of the latent-space dimension}\label{app:pca}

This appendix details the choice of the latent dimension $d$. The intrinsic dimensionality of the thermochemical state space is first estimated with a principal component analysis (PCA) of the input vectors of both the catalytic and non-catalytic min-max scaled datasets $\mathbf{X} \in [0,1]^{N\times D}$, and the resulting choice is then assessed through the output prediction error of the complete reduced-order model.

PCA seeks an orthogonal linear transformation that maximizes the variance captured by successive projections, obtained from the eigendecomposition of the empirical covariance matrix,
\begin{equation}
    \boldsymbol{\Sigma} = \frac{1}{N-1}
    \big(\mathbf{X}-\bar{\mathbf{X}}\big)^\top
    \big(\mathbf{X}-\bar{\mathbf{X}}\big)
    = \mathbf{W}\boldsymbol{\Lambda}\mathbf{W}^\top,
    \label{eq:cov}
\end{equation}
where $\bar{\mathbf{X}}$ denotes the column-wise sample mean of
$\mathbf{X}$, $\boldsymbol{\Lambda} = \mathrm{diag}(\lambda_1, \dots, \lambda_D)$, with $\lambda_1 \geq \lambda_2 \geq \dots \geq \lambda_D \geq 0$, are the eigenvalues of $\boldsymbol{\Sigma}$ and $\mathbf{W} = [\mathbf{w}_1, \dots, \mathbf{w}_D]$ the corresponding orthonormal eigenvectors (or principal components). Each eigenvalue $\lambda_k$ quantifies the variance of the data along its associated principal direction $\mathbf{w}_k$. The cumulative explained variance retained by the first $n$ components is defined as
\begin{equation}
V(n) = \sum_{k=1}^{n} r_k, \qquad n = 1, \dots, D,
\label{eq:cumvar}
\end{equation}
where $r_k=\lambda_k / (\sum_{i=1}^{D} \lambda_i)$ denotes the fraction of the total variance explained by the $k$-th principal component. By construction $V(D) = 1$. We define the minimal number of components $n^\ast$ required to represent the dataset within a prescribed fidelity as
\begin{equation}
    n^\ast = \min \left\{ n \in \{1, \dots, D\} \; : \; V(n) \geq \tau \right\},
    \label{eq:nstar}
\end{equation}
where $\tau = 0.95$ is the retained-variance threshold adopted in this paper.

\begin{figure}[t]
    \centering
    \begin{tabular}{ll}
     (a) & (b) \\
    \includegraphics[scale=0.55,trim=0 5 0 0,clip]{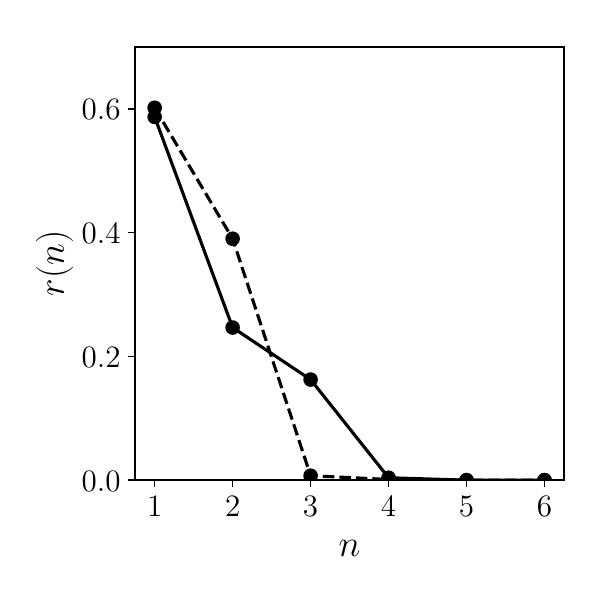} & 
    \includegraphics[scale=0.55,trim=0 5 0 0,clip]{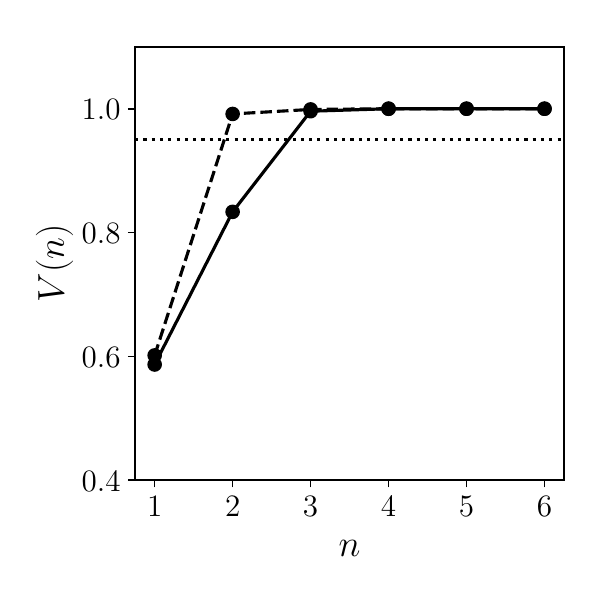}
    \end{tabular}
    \caption{(a) Explained variance ratio $r(n)$ and (b) cumulative explained variance $V(n)$ as a function of the number of retained principal components $n$ for the
    catalytic (solid) and noncatalytic (dashed) datasets. The horizontal dotted line on (b) indicates the $\tau = 0.95$ variance threshold.}
    \label{fig:pca_variance}
\end{figure}

Figure \ref{fig:pca_variance} shows (a) the scree plot, representing the variance ratio of each principal component $r(n)$ and (b) the cumulative explained variance $V(n)$ as a function of the number of principal components $n$ for the catalytic (solid line) and noncatalytic (dashed line) datasets. Looking at the cumulative, in both cases $V(n)$ rises sharply for the first few components and saturates to unity, indicating that the input space, despite being nominally $D$-dimensional, exhibits strong linear redundancy: a small number of principal directions is sufficient to capture nearly all of the variance in the data. This motivates the use of a low-dimensional latent representation. The non catalytic dataset reaches the prescribed variance threshold with two principal components, whereas the catalytic dataset requires three, yielding a cumulative $99.6\%$ at $n^\ast=3$. This indicates that the catalytic data exhibit a higher intrinsic dimensionality than the non catalytic ones. 

\begin{table}[t]
  \centering
  \begin{tabular}{llccc}
    Dataset & $d$ & mean $|e_T|$ (\%) & 99th perc.\ $|e_T|$ (\%) & $\mathrm{NRMSE}_Z$ \\
    \hline
    \multirow{4}{*}{Non-catalytic}
      & 2 & $0.106 \pm 0.012$ & $0.92 \pm 0.26$ & $0.021 \pm 0.001$ \\
      & 3 & $0.020 \pm 0.002$ & $0.12 \pm 0.02$ & $0.007 \pm 0.001$ \\
      & 4 & --$^\dagger$ & --$^\dagger$ & --$^\dagger$ \\
      & 5 & --$^\dagger$ & --$^\dagger$ & --$^\dagger$ \\
    \hline
    \multirow{4}{*}{Catalytic}
      & 2 & $0.515 \pm 0.007$ & $5.38 \pm 0.79$ & $0.107 \pm 0.009$ \\
      & 3 & $0.109 \pm 0.008$ & $0.74 \pm 0.07$ & $0.031 \pm 0.002$ \\
      & 4 & $0.067 \pm 0.006$ & $0.56 \pm 0.05$ & $0.022 \pm 0.002$ \\
      & 5 & $0.040 \pm 0.002$ & $0.27 \pm 0.05$ & $0.011 \pm 0.002$ \\
    \hline
  \end{tabular}
    \caption{Output prediction error on the test set as a function of the latent dimension $d$: mean and 99th percentile of the relative temperature error $|e_T|$ (\%), and root-mean-square error of the thermochemical outputs, normalized by their standard deviation and averaged over all outputs, $\mathrm{NRMSE}_Z$. Mean $\pm$ standard deviation over three trainings with different random initializations. $^\dagger$Ill-conditioned local RBF systems (see text); with a second-degree polynomial term, the mean $|e_T|$ is $0.021\%$ for $d=4$ and $0.024\%$ for $d=5$.}
    \label{tabC:latentdim}
\end{table}

The explained variance of the inputs only provides a heuristic for the latent dimension, since the IO-E is trained to predict the outputs $\mathbf{Z}$. To assess the choice $d=3$ directly, the complete reduced-order model (encoder, community detection and local RBF surrogates) was trained for $d=2,3,4$ and $5$, with all other parameters unchanged (Appendix~\ref{app:ae_parameters}), and evaluated on the held-out test set. Each model was trained three times with different random initializations. Table~\ref{tabC:latentdim} reports the mean and 99th percentile of the relative temperature error and the root-mean-square error of the outputs, normalized by their standard deviation and averaged over all outputs. For the catalytic dataset, the error decreases sharply from
$d=2$ to $d=3$, by a factor of 3.5 to 7 for all metrics, confirming that two latent dimensions do not suffice to represent the catalytic states. Beyond $d=3$ the error decreases only gradually, from a mean temperature error of
about $0.1\%$. For the non-catalytic dataset, $d=3$ yields the lowest error. For $d \ge 4$, the latent coordinates remain nearly three-dimensional, and the polynomial term of the local RBF networks becomes ill-conditioned; when these networks are refitted with a second-degree polynomial, the error returns to the level obtained with $d=3$, so that the additional dimensions bring no improvement. The latent dimension $d=3$ is therefore retained for both datasets, as the smallest dimension beyond which the error decreases only gradually, while keeping the cost of the local surrogates low and the latent manifold directly
interpretable.

\section{Input–Output encoder architecture and training parameters}
\label{app:ae_parameters} 

\begin{table}[htbp]
\centering
\begin{tabular}{ll}

Parameter & Value \\
\hline
Encoder architecture & Fully connected: $64 \rightarrow 32 \rightarrow d$ \\
Decoder architecture & Fully connected: $d \rightarrow 32 \rightarrow 64 \rightarrow D_z$ \\
Hidden-layer activation & $\tanh$ \\
Latent activation & Linear \\
Output activation & Linear \\
Latent dimension & $d=3$ \\
Mini-batch size & $N_b=256$ \\
Optimizer & Adam (default hyper-parameters) \\
Training epochs & $2000$ \\
Stress weight & $\lambda_{\mathrm{stress}}=1$ \\
Loss function & Eq.~\eqref{eq:ioe_loss} \\
\hline
\end{tabular}
\caption{Architecture and training hyper-parameters of the input--output encoder (IO-E).}
\label{tab:ae_parameters}
\end{table}

The implementation details of the input-output encoder used for the manifold learning analysis are summarized in Table~\ref{tab:ae_parameters}. Specifically, the encoder architecture consists of two fully connected hidden layers of $64$ and $32$ neurons with
hyperbolic-tangent ($\tanh$) activation functions, followed by a linear bottleneck layer
of width $d$. The decoder has the mirrored
architecture ($32$, then $64$ neurons, $\tanh$ activations, linear
output layer). The cost functional in Eq.~\eqref{eq:ioe_loss} is evaluated using mini-batches of size $N_b=256$, avoiding the prohibitive computational cost associated with the $\mathcal{O}(N^2)$ pairwise-distance calculations required by the loss function when applied to the full dataset. The network is trained for $2000$ epochs with the Adam optimizer, with its default hyper-parameters (learning rate $\eta_{r}=10^{-3}$, exponential decay rates $\beta_1=0.9$ and $\beta_2=0.999$, and numerical stabilization parameter $\epsilon_{\mathrm{A}}=10^{-7}$). The aforementioned set of parameters is adopted for both the catalytic and noncatalytic datasets.


\end{appendices}

\bibliography{sn-bibliography}

\end{document}